\documentclass[12pt]{article}
\usepackage{amsfonts}
\usepackage{amssymb}
\usepackage{graphicx}
\usepackage{color}

\newcommand {\exe} {\stackrel{\cdot} {=}}
\newcommand {\reals} {{\rm I\!R}}

\newcommand {\bx} {\mbox{\boldmath $x$}}

\newcommand {\bE} {\mbox{\boldmath $E$}}

\newcommand {\bX} {\mbox{\boldmath $X$}}

\newcommand{\calH}{{\cal H}}

\newcommand{\calN}{{\cal N}}

\newcommand{\calP}{{\cal P}}

\newcommand{\calX}{{\cal X}}

\begin{document}

\title{Statistical Physics of Coding for the Integers}

\author{Neri Merhav\\[1ex]
\small The Andrew \& Erna Viterbi Faculty of Electrical Engineering\\
\small Technion -- Israel Institute of Technology, Haifa 32000, Israel\\
\small E-mail: merhav@ee.technion.ac.il}

\date{}

\maketitle

\begin{abstract}
We consider coding of the positive integers under the zeta distribution
$P_\beta(x)\propto x^{-\beta}$, whose normalizing constant is the Riemann zeta
function $\zeta(\beta)$. With the idealized Hamiltonian $\calH(x)=\ln x$, which
saturates the Kraft--McMillan lower bound on code lengths, this problem
corresponds to the statistical mechanics of a Hagedorn system: the density of
states grows exponentially with energy, so the partition function $\zeta(\beta)$
converges only for $\beta>1$. That $\zeta(\beta)$ is a Hagedorn-type partition function is
known from the primon-gas literature; the new connection here is the
observation that the Kraft--McMillan counting argument motivates the idealized
Hamiltonian whose density of states grows exponentially, which opens the 
way to a number of new results. We derive the micro-canonical entropy with an
Euler--Mascheroni sub-leading correction, the critical exponent of the energy
fluctuations at the Hagedorn temperature, and full equivalence of the
canonical and micro-canonical ensembles; the large-deviations rate function
for mismatched coding, exactly a Kullback--Leibler divergence within the
zeta family (above the mean energy); and the resulting buffer-overflow threshold. The
critical exponent is the Fisher information of the family, which is the natural
reason for a singularity of universal coding at the Hagedorn point: we prove directly
that the minimax regret and the minimax expected redundancy of the zeta
family are infinite when the parameter range reaches that point, whereas staying a
margin $\delta$ away from it costs only a doubly-logarithmic penalty in the
second-order term as $N\to\infty$. Finally, we give an explicit
two-part integer code (a Golomb code for the dyadic scale of the integer, followed
by raw offset bits, with a single integer parameter) whose expected redundancy is provably bounded up to the
critical point -- numerically, about $0.06$--$0.07$ bits at the tabulated values 
$\beta-1\le0.1$ -- whereas the standard integer codes (Elias $\gamma$, $\delta$ 
and $\omega$, and the $\zeta_k$ codes of Boldi and Vigna)
have much larger absolute redundancy there, which grows as $\beta\to1^+$ 
(numerically over the computed range $\beta\le1.5$, and provably for every $\zeta_k$, Elias $\gamma$ included).
The code is tailored to $\beta$ through its single integer parameter (for 
which the explicit choice $m=\lceil1/(\beta-1)\rceil$ already gives the bound), 
and is not universal over the family, which Sec.~5 shows to be impossible near $\beta=1$.
\end{abstract}


\section{Introduction}

One of the basic problems in information theory is the assignment 
of code lengths to elements of a countable set for the purpose of data
compression, in particular, compact representation for all the positive
integers by the shortest possible binary strings.
Coding schemes for the integers play a central role in a variety of settings where the objects to be described are drawn 
from a countably infinite or a priori unknown domain. 
This situation arises ubiquitously in data compression, where algorithms such
as the Lempel-Ziv algorithm in its various versions (LZ77 \cite{ZL77}, LZ78
\cite{ZL78} and others)
must encode dictionary indices, match lengths, and offsets that grow with the data, making fixed-length representations infeasible. More broadly, 
integer coding underlies universal modeling and inference: in the minimum
description length (MDL) principle (see, e.g., \cite{Rissanen83} and
references therein), model classes are penalized via the code-lengths of
integer-valued parameters (Markov order, number of states, etc.), 
while in Kolmogorov complexity, self-delimiting descriptions require efficient encodings of lengths and indices. 
Practical systems likewise rely on compact representations of counts, gaps, and run-lengths in compressed indexes and streaming protocols. 
From a probabilistic perspective, integer codes are also used for heavy-tailed distributions 
(e.g., Zipf-like laws to be discussed below), thereby bridging combinatorial structure and statistical modeling. 
Collectively, these applications highlight that efficient, prefix-free coding of the integers 
is not merely a technical convenience but a fundamental building block in information theory, compression, and learning.

The starting point of this paper is a connection, which to the best of our
knowledge is new, between two fields that rarely meet: the Kraft--McMillan
counting constraint on code lengths for the integers, and the statistical
mechanics of a Hagedorn system, i.e., a model whose density of states grows
exponentially with energy, so that its canonical partition function has a
finite abscissa of convergence beyond which the usual thermodynamic description
breaks down. For the idealized Hamiltonian $\calH(x)=\ln x$, the ``energy'' of
an integer $x$ is simply (proportional to) its ideal code length, and, in
this idealized setting, the fact that the Kraft--McMillan constraint permits
exponentially many integers up to a given code length is what makes the
exponential growth of the density of states, and hence Hagedorn behavior, natural for this Hamiltonian. To be precise about
what is and is not claimed: that the zeta function $\zeta(\beta)=\sum_{x=1}^\infty x^{-\beta}$,
$\beta>1$ -- the normalizing constant of the zeta (Zipf) law $P_\beta(x)=x^{-\beta}/\zeta(\beta)$ on the positive
integers $x=1,2,\ldots$, with $\beta$ playing the role of an inverse temperature -- is a
Hagedorn-type partition function is known from the physics literature on
arithmetic statistical mechanics (the primon/Riemann gas), discussed and cited
explicitly in Sec.~3.1.1 below. What we are not aware of in that literature
is the interpretation of this partition function as that of a \emph{coding}
problem, with the energy as code length and the Kraft--McMillan counting
argument as the motivation for the idealized exponential growth, and hence the possibility
of importing the tools of coding and information theory into the study of the
Hagedorn system, and vice versa. We lay out this connection in a self-contained
way, in the hope that it will be useful on both sides of the divide.

Beyond the connection itself, the paper's contribution is a set of new
results that it motivates. They answer a specific question: once the integers are
equipped with the zeta/Zipf law $P_\beta$, how closely -- and how
\emph{uniformly} as $\beta\downarrow1$ -- can the Shannon bound be approached
by a single explicit coding scheme, what happens to the large-deviations
behavior of mismatched coding as $\beta$ approaches this point, and what does
this cost in terms of the achievable minimax redundancy of universal coding
across the whole family? The answers, five in number, are new to the best of
our knowledge. Each is stated as an explicit, checkable result tied to a
specific section.
\begin{enumerate}
\item \emph{A coding scheme whose redundancy stays bounded all the way to the
critical point} (Sec.~6). The idea follows from the heavy-tail picture of Secs.~2--5: near $\beta=1$, a power law in $x$ becomes a nearly flat geometric law in the dyadic scale $\lfloor\log_2x\rfloor$, which calls for a Golomb code rather than a unary one, while within a dyadic block the law is nearly uniform. We construct a single, explicit, two-part
integer code -- a Golomb code for the dyadic scale of the integer, with the
Golomb parameter chosen to minimize the expected length, followed by the raw
offset bits within the dyadic block -- that applies to
the zeta law with any $\beta>1$. An exact entropy-bookkeeping identity
(not an approximation) decomposes its expected redundancy into two terms, each of
which we bound by an explicit constant: the redundancy is provably below $3.7$ bits for all $1<\beta\le2$, and its limit superior as $\beta\to1^+$ is at most
$2.17$ bits (Proposition~1), and exact computation
shows it to be much smaller: at the values of $\beta$ tabulated in Table~\ref{tab:redundancy}, from $\beta=3$ down to $\beta=1.001$,
it lies between $0.058$ and $0.443$ bits.

The code coincides with Elias $\gamma$ coding when the Golomb parameter is $m=1$ and with
the $\zeta_2$ code of Boldi and Vigna \cite{BoldiVigna05} (of which Elias $\gamma$ is the case $k=1$) when $m=2$. Near criticality it separates from the $\zeta_k$ codes:
we prove (Proposition~2) that the redundancy of the best $\zeta_k$ grows without bound as $\beta\to1^+$, even when $k$ is
re-optimized at every $\beta$ -- numerically, about $43$ bits at $\beta-1=0.001$ for the
best $\zeta_k$ against $0.058$ bits for the present code. The code is tailored to $\beta$ through its single integer parameter (the explicit choice $m=\lceil1/(\beta-1)\rceil$ already gives the bound of Proposition~1)
and, as Sec.~5 shows must be the case, is not universal over the family.
\item \emph{Universal coding is singular exactly at the Hagedorn point, with
a quantified price for staying away from it} (Sec.~5). The Fisher
information of the zeta family is the variance of the energy, which diverges
as $(\beta-1)^{-2}$; this is the natural reason for a singularity of universal
coding at $\beta=1$, but, since divergence of the Jeffreys integral alone does not
imply infinite redundancy (Sec.~5), we prove directly that minimax universal coding of the family has infinite
redundancy when the parameter range reaches $\beta=1$. We prove this exactly,
not heuristically: the one-sample Shtarkov sum diverges
($M(x)\sim1/[ex(\ln x+\gamma)]$), which gives infinite minimax regret; a separate short
argument shows that the minimax expected redundancy is infinite too; and both conclusions extend to every sample
size $N$ (for the regret, since a supremum of a sum is at most the sum of the suprema; for the expected redundancy, by data processing). 

Staying a margin $\delta$ away from criticality
costs only a \emph{doubly-logarithmic} penalty,
$O(\log\log(1/\delta))$, in the second-order term of the minimax expected redundancy as $N\to\infty$ for each fixed $\delta$: going from $\delta=0.1$ to $\delta=10^{-6}$ adds only
about $1.4$ nats. 

Thus the singularity of the physical model is reflected sharply in universal
coding: the redundancy is infinite when the parameter range reaches $\beta=1$ and, for a fixed margin and $N\to\infty$,
grows only doubly-logarithmically as the margin shrinks.
\item \emph{A critical exponent and a sub-leading correction for an
explicitly solvable Hagedorn system} (Sec.~3). We derive a sub-leading,
Euler--Mascheroni-constant correction to the leading-order Hagedorn entropy,
together with the associated critical exponent
$\mathrm{Var}_\beta(\ln X)\sim(\beta-1)^{-2}$ (variance under $P_\beta$) for
the divergence of the energy fluctuations at the critical point. This
exponent is not merely a thermodynamic curiosity: it is the Fisher
information of the family, and it is the natural reason for the singularity of universal
coding established in Sec.~5.
\item \emph{Equivalence of the canonical and micro-canonical ensembles,
proved} (Sec.~3). We give a complete proof, from an independently defined
micro-canonical entropy (defined by directly counting $N$-tuples whose total energy lies in a thin
shell, not merely the Legendre transform taken as a definition), that the two ensembles are equivalent, at the level of
thermodynamic functions, on the entire finite-energy domain, with the Hagedorn temperature
appearing only as a limiting, never-attained temperature. Equivalence of
ensembles for a Hagedorn system cannot simply be taken for granted by analogy with systems where it is known to hold; here it is
proved for this model.
\item \emph{Large deviations of mismatched coding for the zeta family} (Sec.~4).
The large-deviations rate function of the total code length is shown to be
exactly a Kullback--Leibler divergence within the zeta family (for $\epsilon$ above the mean energy),
$I_\beta(\epsilon)=D\bigl(P_{\tau^*(\epsilon)}\|P_\beta\bigr)$, and the
buffer-overflow problem is worked out explicitly for this family: the optimal mismatched coding
parameter $\theta^*$ satisfies $H(P_{\theta^*})=R\ln2$, so that it depends on
the buffer rate $R$ alone and not on the unknown $\beta$ -- a code designer
needs to know only the buffer size -- and the overflow exponent is positive
if and only if $R>H(P_\beta)/\ln2$.
\end{enumerate}
These results do not follow by simply invoking the primon/Riemann-gas
analogy; each requires an explicit derivation for the zeta family. Taken
together, they turn an analogy into a working toolkit: a code, a limit on
universal coding, and a tractable thermodynamic model, all tied to the
same critical point. We note, for context, that the general translation
between Zipf-law statistics and Hagedorn-type linear entropy-energy scaling
has also surfaced independently in an unrelated setting, namely as a
description of human brain activity from fMRI data
\cite{ChialvoJanik24}; this suggests that the translation may arise
whenever a power-law distribution is read as a Boltzmann
distribution, of which the coding-theoretic instance developed here is a
more fully worked-out example.

Consider any uniquely decodable code that assigns code lengths $\ell(x)$ to the positive integers, 
$x=1,2,\ldots$, arranged in non-decreasing order, $\ell(1)\le\ell(2)\le\ell(3)\le \cdots$.
As will be shown below, in order for
the code for the natural numbers to be uniquely decodable,
the code-length for the integer $x$ is lower bounded by
\begin{equation}
\label{lowerbound}
\ell(x) \ge \log x.
\end{equation}
Thus, even in the absence of any probabilistic assumptions, there is an intrinsic logarithmic 
growth of code length with the index, which cannot be avoided. This reflects a basic combinatorial constraint: 
describing larger and larger objects necessarily requires increasing resources.

A natural question which may arise at this point is whether this lower bound can be approached 
in a universal manner, without assuming any specific probability distribution over the integers. 
A classical answer was provided by Elias \cite{Elias75} who proposed several
simple and efficient coding schemes
which assign to each positive integer a binary representation; for the $\delta$ and $\omega$ codes the length behaves as
\begin{equation}
\ell(x) = \log x + O(\log \log x).
\end{equation}
These constructions achieve, up to lower-order corrections, the minimal growth
compatible with the lower bound (\ref{lowerbound}), at least asymptotically for large
$x$. In this sense, they realize a nearly optimal way of encoding the integers in a distribution-free setting.

Once a probabilistic structure is introduced, the problem acquires an additional layer. 
If the integers are generated according to a given probability distribution,
$\{P(x),~x=1,2,\ldots\}$, then neglecting integer length constraints, the optimal code length is given by $\ell(x)
= -\log P(x)$ \cite{CT06}. With this fact in mind, it is readily
observed that Elias $\delta$ and $\omega$ coding for the integers corresponds to a probability
assignment of the form
\begin{equation}
P(x)= \frac{1}{2^{\log x+O(\log\log x)}}=\frac{1}{x\cdot 2^{O(\log\log x)}},
\end{equation}
namely, a distribution that decays at a rate that is slightly faster than the
rate of $\frac{1}{x}$ (see also \cite{Rissanen83}). But this extra speed of
convergence beyond $\frac{1}{x}$ is
inevitable since the sum of the infinite series $\sum_{x=1}^\infty
\frac{1}{x}$ diverges, and so, a distribution proportional to
$\frac{1}{x}$ cannot exist, unless truncated to a finite
range, $x\in\{1,2,\ldots,J\}$. A simple natural remedy, in the case of infinite
support, is to introduce a power parameter $\beta>1$, namely, to let $P(x)$ be
proportional to $\frac{1}{x^\beta}$, whose sum is convergent, as will be
discussed shortly.

Interestingly, distributions of this type are not merely of theoretical interest.
They are closely related to empirical regularities such as Zipf's law (see,
e.g., \cite{Powers98}, \cite{Piantadosi14} and references therein), according to which the
frequency of an event is inversely proportional to its rank, and to its refinement,
the Zipf--Mandelbrot law (see, e.g., \cite{Mandelbrot68}), 
which incorporates a shift parameter and possibly a power parameter, that is,
\begin{equation}
\label{mandelbrot}
P(x) \propto \frac{1}{(x+\alpha)^\beta},~~~~\alpha\ge 0,~\beta>1.
\end{equation}
Such probability laws have been empirically observed
across a wide range of systems, including natural language, biological data, and complex networks.
From the coding perspective,
they correspond to situations in which large integers occur with non-negligible
probability, making logarithmic-length codes not only necessary but essentially optimal (to leading order).

This class of heavy tailed probability distributions is referred to as {\em
power law distributions}. 
A particularly important subclass of (\ref{mandelbrot}) is given by the
special case of $\alpha=0$, namely,
\begin{equation}
P_\beta(x)= \frac{x^{-\beta}}{\zeta(\beta)}, \qquad \beta>1,~~x=1,2,\ldots
\end{equation}
where $\zeta(\beta)$ is the normalization constant,
\begin{equation}
\label{zetafunction}
\zeta(\beta)=\sum_{x=1}^\infty \frac{1}{x^\beta},
\end{equation}
which is the well known {\em Riemann zeta function}. This leads to
code lengths of the form $\ell(x) \approx \beta \log x$ (for large $x$), which respect the fundamental logarithmic 
scaling dictated by counting, while introducing a parameter that controls the
relative weight assigned to large integers.
Baer \cite{Baer07}, \cite{Baer08} suggested structured, efficient algorithms
of coding for the zeta distribution. In Sec.~6, we propose an alternative
algorithm: an explicit two-part rule with a single integer parameter.
In this context, it should be noted that the Elias $\delta$ and $\omega$ codes can be viewed as corresponding to a choice of $\beta$ that
depends on $x$, in particular, $\beta_x=1+O(\log(\log x)/\log x)$.

The normalization of the power-law distribution, which is the zeta function,
$\zeta(\beta)$, can be naturally displayed as a partition function,
\begin{equation}
\label{partitionfunction}
\zeta(\beta) = \sum_{x=1}^\infty e^{-\beta\ln x}.
\end{equation}
Interpreting $\ln x$ as a Hamiltonian, $\calH(x)$ associated with state $x$, this defines a statistical-mechanical model 
with an unbounded state space. According to this model, the energy,
$\calH(x)=\ln x$, associated with state $x$ is
proportional to the cost of its code length,
and coding with respect to the distribution $P_\beta$
corresponds to the canonical ensemble associated with $\calH(x)$.
The divergence of $\zeta(\beta)$ at 
$\beta=\beta_{\mbox{\tiny c}}=1$ signals a critical point at 
which the normalization breaks down, reflecting the overwhelming contribution of high-energy states.

In more detail, the above described statistical-mechanical
model is analogous to a Hagedorn system \cite{Hagedorn65},
\cite{AW88}, \cite{CP75},
which is a system whose
density of states grows exponentially with the energy, up to a possible
sub-exponential multiplicative factor (see Appendix B for more detailed
background). The above mentioned critical point
$\beta_{\mbox{\tiny c}}=1$, at and below which $\zeta(\beta)$ diverges, is exactly the Hagedorn inverse temperature
associated with this model. 

It is important to emphasize that this phenomenon is not a consequence
of any specific interaction or ``potential'' in the usual sense, but rather of the relation between the indexing of states
and their assigned energy. In our setting, the energy variable is naturally identified with the code length, which,
in accordance with the fundamental counting constraint, grows logarithmically
with the integer label, $\calH(x) = \ln x$. The exact counting function (with $\calN=\{1,2,3,\ldots\}$ the positive integers),
\begin{equation}
\label{countingfunction}
N(E):=|\{x\in\calN:\calH(x)\le E\}|=\lfloor e^E\rfloor,
\qquad E\ge0,
\end{equation}
is then known exactly, with no approximation. Writing $\lfloor y\rfloor=y-\{y\}$
($\{y\}\in[0,1)$ the fractional part), the number of states in any window
$(E,E+\Delta]$ satisfies the exact identity
\begin{equation}
\label{windowcount}
N(E+\Delta)-N(E)=e^E(e^\Delta-1)+r(E,\Delta),\qquad |r(E,\Delta)|<1,
\end{equation}
valid for every $E\ge0$ and $\Delta>0$, with no discretization or limiting
argument involved. In particular, for any \emph{fixed} $\Delta>0$, dividing
by $e^E$ and letting $E\to\infty$ makes the remainder vanish,
$[N(E+\Delta)-N(E)]/e^E\to e^\Delta-1$; only then, in a second and separate
step, letting $\Delta\to0$ (so that $(e^\Delta-1)/\Delta\to1$) recovers the
density of states in the usual continuum sense,
\begin{equation}
\label{densityofstates}
\hat\rho:=\lim_{\Delta\to0}\ \lim_{E\to\infty}\frac{N(E+\Delta)-N(E)}{\Delta\, e^E}=1,
\end{equation}
i.e., the density of states itself grows as $e^{\beta_{\mbox{\tiny c}}E}$,
with the two limits kept explicitly separate and taken in this order --
$E\to\infty$ at fixed $\Delta$, only afterward $\Delta\to0$ -- which is what
makes the exponential growth of the density of states a rigorously
established fact rather than an informal one.
This exponential proliferation of states
is precisely the feature underlying Hagedorn behavior, leading to a finite radius of convergence of the partition function
and a critical point at which normalization breaks down. In other words, in contrast to more traditional settings,
where such growth arises from intricate microscopic structure,
here it is a direct consequence of the logarithmic scaling of the idealized Hamiltonian $\calH(x)=\ln x$, suggested by coding considerations.
This provides a particularly transparent realization of a Hagedorn transition, rooted in the combinatorial structure of the integers.

Moreover, owing to Euler's product form of the zeta function, there is also another
analogy, pertaining to the grand-canonical ensemble of the Bose gas, whose energy levels are $E_p
=\ln p$, where $p$ runs over all primes, i.e., $p=2,3,5,\ldots$. As $\beta$
descends towards $\beta_{\mbox{\tiny c}}=1$, the total number of bosons in this
model goes to infinity. 

All the above holds true even for a single `particle' at state $x$. But to
complete the picture, we also
consider a system with $N$ independent such particles, $(x_1,\ldots,x_N)$,
and focus on its density of states for large $N$, showing that for large
per-particle energy $\epsilon$, it behaves roughly
like $\exp\{N\beta_{\mbox{\tiny c}}\epsilon\}$ (where again, $\beta_{\mbox{\tiny
c}}=1$), and so, the Hagedorn system characteristics are of course manifested here too.
It will be interesting to note that, although the canonical parameter $\beta$
is necessarily restricted to $\beta>\beta_{\mbox{\tiny c}}=1$ for the partition
function to converge in the first place, the canonical and micro-canonical
descriptions remain fully equivalent throughout this entire domain: the map
from $\beta$ to the mean energy per particle is a strictly monotonic bijection
onto the whole positive real line, so every finite energy is reached by a
unique inverse temperature. The only singular feature is a limiting
(Hagedorn) temperature, $T_{\mbox{\tiny c}}=1/\beta_{\mbox{\tiny c}}=1$, that is
approached, but never attained at any finite energy, as the energy per
particle diverges. By contrast, if the support of $P_\beta$ is truncated into
a finite range, $x\in\{1,2,\ldots,J\}$, the partition function converges for
every real $\beta$ (positive or negative), the micro-canonical entropy attains its maximum $\ln J$ at the uniform-distribution energy and decreases to zero at the top of the finite energy range, $\epsilon=\ln J$, and there is no longer any
limiting temperature.

Finally, from the data compression perspective, 
we also study the behavior of code lengths for sequences of independent samples. 
Our focus is on the large deviations properties of the total code length and on the interplay 
between the coding parameter and the statistics of rare events. We show that the optimal parameters 
governing these deviations are driven toward the critical point
$\beta_{\mbox{\tiny c}}=1$ as the deviation level grows, leading to asymptotic behavior analogous to 
that of systems with Hagedorn-type transitions. 

These results provide a concrete setting in which fundamental ideas from
information theory -- universal coding for natural numbers, heavy-tailed distributions, and large deviations -- interact 
with concepts from statistical mechanics such as partition functions, criticality, and ensemble equivalence.

\section{Background on Uniquely Decodable Data Compression Codes}

A code for lossless data compression is a mapping between a finite or
countable alphabet of symbols of the data to be compressed (e.g., latin letters,
digits, or any other characters) into a set of binary strings,
which may be of different lengths. For the data to be perfectly recoverable
from its compressed representation, this mapping must be one-to-one, not only
in the level of single symbols and their codewords, but also when the code is
used repeatedly and the decoder receives the concatenation of the compressed
binary strings corresponding to the successive data symbols, because the parsing of
the codewords is not provided to the decoder. Throughout, code lengths
$\ell(x)$ are measured in bits, i.e., using $\log=\log_2$, whereas the
Hamiltonian $\calH(x)=\ln x$ and the partition function $\zeta(\beta)$ in
Sec.~3 onward are expressed in nats, i.e., using the natural logarithm; this
is purely a matter of units and does not affect the location of the critical
point $\beta_{\mbox{\tiny c}}=1$, which is a property of the power series
$\sum_x x^{-\beta}$ itself, independent of the base used to express energies.

A simple method to ensure this
unique decodability property is to design the code by maintaining the
prefix-free condition, namely, to keep the rule that no codeword would be a prefix of any
other codeword. For example, if the alphabet is
$\calX=\{\mbox{A},\mbox{B},\mbox{C}\}$ and the corresponding codewords are
`0', `10', and `11', the prefix condition holds, and there is only one way
(and hence - the right way) 
to parse any given compressed bit-stream and thereby to decode it. But it is not necessary to maintain the
prefix condition for the code to be uniquely decodable (UD), as there are UD
codes which are not prefix-free.
Whether the
prefix condition is met or not, it is clear that a necessary condition for
a code to be UD is that the codewords of the code are sufficiently long in
some collective sense. 
To give this (admittedly vague) statement a clear significance, let us first define the {\em length
function}, $\ell(x)$ ($x\in\calX$) of the given code to designate the length
(in bits) of the codeword for $x$. In the above example, $\ell(\mbox{A})=1$,
and $\ell(\mbox{B})=\ell(\mbox{C})=2$.
One of the fundamental principles in information theory is a necessary
and sufficient condition for the existence of a UD code with a given length
function, $\ell(x)$. This necessary and sufficient condition 
is the well-known Kraft-McMillan (KM) inequality (see, e.g., \cite{CT06}):
\begin{equation}
\sum_{x\in\calX} 2^{-\ell(x)}\le 1.
\end{equation}
In the above example,
\begin{equation}
2^{-\ell(\mbox{\tiny A})}+
2^{-\ell(\mbox{\tiny B})}+
2^{-\ell(\mbox{\tiny C})}=2^{-1}+2^{-2}+2^{-2}=1,
\end{equation}
and so, the KM inequality is saturated in this case.
Note that whenever the KM inequality is met with equality,
the terms of the KM sum, $\{2^{-\ell(x)},~x\in\calX\}$ can be thought of as
probabilities, $\{P(x),~x\in\calX\}$, as they are non-negative reals that sum to
unity. Conversely, every probability distribution with dyadic probabilities
(integer powers of $\frac{1}{2}$) correspond to a length function,
$\ell(x)=-\log P(x)$ of some UD code. Thus,
lossless compression and probability assignment are two sides
of the same coin.

Consider now the case where the alphabet is $\calX=\calN\equiv\{1,2,3,\ldots\}$,
namely, the set of all natural numbers, and we wish to design a data
compression code for $\calN$. It is conceivable that large natural numbers
would need to be mapped to longer codewords than those of small natural
numbers, and therefore, $\ell(1)\le\ell(2)\le\ell(3)\le\cdots$.\footnote{Even if this
is not the case, one can always rearrange the elements of $\calX$ according to
non-decreasing values of $\ell(x)$ and relabel the members of this alphabet.}
As observed in \cite{Rissanen83}, the KM inequality and the monotonicity of
$\ell(x)$ imply the following:
\begin{equation}
1\ge \sum_{x'=1}^\infty 2^{-\ell(x')}
\ge\sum_{x'=1}^x 2^{-\ell(x')}\ge x\cdot 2^{-\ell(x)},
\end{equation}
and so,
\begin{equation}
\label{lowerbound2}
\ell(x)\ge\log x.
\end{equation}
In other words, no UD code can assign to any $x\in\calX$ a codeword shorter than the base 2
logarithm of the rank of $x$ in the order of non-decreasing $\ell(x)$. This is a fundamental limitation of data
compression, that stems from
purely combinatorial considerations, regardless of any probability
distribution that may govern the data, and even in the absence of any
probabilistic model. Therefore, any concrete coding scheme
whose length function comes close to $\log x$ is optimal or nearly optimal to leading order.

As mentioned in the Introduction, the Elias $\delta$ and $\omega$ codes \cite{Elias75}
come close as they have length functions of the form
$\ell(x)=\log x+O(\log(\log x))$ (the $\gamma$ code, with $\ell(x)=2\lfloor\log_2x\rfloor+1$, is within a factor of $2$), and so, they are nearly optimal at least for large
$x$. Obviously, there is no UD code that meets the lower bound
(\ref{lowerbound2}) for all $x$ since $\sum_{x=1}^\infty 2^{-\log
x}=\sum_{x=1}^\infty\frac{1}{x}=\infty$. In other words, the corresponding KM sum
is not only larger than unity, but it diverges altogether.

Accordingly, considering the above-mentioned correspondence between compression and probability 
assignment, there is no probability distribution, $P(x)$, that is proportional to
$2^{-\log x}=\frac{1}{x}$ for all $x\in\calN$, since this series diverges and
therefore it is not normalizable. 

This divergence is worth pausing on, since it
marks precisely the distinction between the idealized Hamiltonian used
below and an actual, realizable code. The energy function
$\calH(x)=\ln x$ that will play the role of a Hamiltonian in the
statistical-mechanical model of Sec.~3 is the idealized, lower-envelope
Hamiltonian that saturates (\ref{lowerbound2}) with equality for every $x$ --
it is not literally the length function of any one, concretely realizable,
uniquely decodable code. 

Indeed, for the length function $\ell(x)$ of any
actual prefix code, the Kraft--McMillan inequality guarantees
$\sum_x2^{-\ell(x)}\le1$: a KM sum that stays finite, in fact bounded by $1$,
at the ``inverse temperature'' $\beta=1$ for every realizable code. It is
only the idealized Hamiltonian $\calH(x)=\ln x$ -- which no single code can
realize exactly, for every $x$, since that would require the divergent KM
sum just exhibited -- whose associated partition function $\zeta(\beta)$
diverges at $\beta=1$, precisely the Hagedorn critical point at the center
of this paper. The correspondence developed in what follows should be
understood in this sense: a statement about the idealized combinatorial
Hamiltonian that saturates the universal lower bound on code lengths, not
a claim that any particular realizable code's own partition function
diverges at $\beta=1$ (to the contrary, it cannot, by Kraft--McMillan). We
do not claim, conversely, that every other thermodynamic quantity of an
actual code's partition function must stay bounded at $\beta=1$: Kraft--McMillan
controls the partition sum itself, not its energy moments, and depending on
the sub-leading growth of a specific code's length function $\ell(x)$, its
mean energy or energy variance may still diverge there.

A natural compromise that comes to mind is then
to resort to a probability distribution that is quite similar, namely, one
where 
\begin{equation}
P(x)\propto\frac{1}{x^\beta},~~~~~\beta>1
\end{equation}
which is the so-called zeta distribution:
\begin{equation}
P(x)=\frac{1}{\zeta(\beta)x^\beta},
\end{equation}
where the normalization constant
is recognized as the Riemann zeta function (\ref{zetafunction}).
We henceforth denote the zeta distribution by $P_\beta(x)$ instead of the
generic notation $P(x)$, that is,
\begin{equation}
P_\beta(x)=\frac{x^{-\beta}}{\zeta(\beta)},
\end{equation}
in order to maintain the dependence on $\beta$ in the notation.
Throughout, $X$ denotes a random variable drawn from $P_\beta$, and
$\bE_\beta\{\cdot\}$ and $\mbox{Var}_\beta\{\cdot\}$ denote expectation and variance
under $P_\beta$ (similarly, $\bE_\theta$, $\bE_{\beta^*}$, etc., under $P_\theta$, $P_{\beta^*}$, etc.).
Here the parameter $\beta$ can be used as a regularization parameter that
controls the degree of proximity to
$\frac{1}{x}$ and the closely related Zipf distribution, the Zipf-Mandelbrot
distribution, etc., as discussed in the Introduction.
Ignoring integer length constraints, the length function pertaining to the zeta distribution
is given by
\begin{equation}
\ell_\beta(x)=-\log P_\beta(x)=\beta \log x+\log\zeta(\beta).
\end{equation}
When $\beta$ comes close to 1 (from above), the ratio $\ell_\beta(x)/\log x$ comes close to 1 as long as
$\log x \gg\log\zeta(\beta)$, although the excess $(\beta-1)\log x+\log\zeta(\beta)$ over the lower bound $\log x$ is unbounded in $x$.
As mentioned earlier, in Sec.~6, we propose a simple coding scheme whose
expected length is within a bounded additive gap of the entropy, i.e., of the expected length of $\ell_\beta(x)$ (Proposition~1).
The optimal choice of $\beta$ is associated with a compromise since the
first term of $\ell_\beta(x)$ is increasing with $\beta$, whereas the second
term is decreasing. If the expectation $\mu$ of $\log X$ is known, or can be
estimated empirically from past data, then the optimal choice of $\beta$ is readily
seen to be the one
for which the expectation of $\log X$ under $P_\beta$ is equal to $\mu$.

\section{The Zeta Partition Function and its Hagedorn
Behavior}

\subsection{Single Particle}

We now focus on the normalization constant of $P_\beta(\cdot)$, which is
$\zeta(\beta)$, and view it as a partition function in the form of eq.\
(\ref{partitionfunction}). Let us examine the tail of the zeta function,
$\zeta_n(\beta)\equiv\sum_{x=n}^\infty x^{-\beta}$ for a given $n\ge 1$, which
constitutes the contribution of high energies (large $E=\ln x$) to the
partition function $\zeta(\beta)$. Since $x^{-\beta}=e^{-\beta\ln x}$ is
monotonically decreasing in $x$, a direct comparison with the integral
$\int_n^\infty u^{-\beta}\mbox{d}u=n^{1-\beta}/(\beta-1)$ (for $\beta>1$)
gives, for every $n\ge 1$,
\begin{equation}
\label{tailbound}
\frac{n^{1-\beta}}{\beta-1}\le\zeta_n(\beta)\le\frac{n^{1-\beta}}{\beta-1}+n^{-\beta},
\qquad\beta>1,
\end{equation}
while $\zeta_n(\beta)=\sum_{x=n}^\infty x^{-\beta}=\infty$ for every $\beta\le
1$. Both bounds in (\ref{tailbound}) are of the same exponential order in
$E=\ln n$, namely,
\begin{equation}
\zeta_n(\beta) \exe \frac{n^{1-\beta}}{\beta-1}=\frac{1}{\beta-1}\,e^{-(\beta-1)E},
\qquad E=\ln n,
\end{equation}
in the sense that the ratio between $\zeta_n(\beta)$ and this expression tends
to 1 as $n\to\infty$ (for fixed $\beta>1$). Equivalently, recalling the
rigorously established density of states $\hat\rho(E)\sim e^{E}$ for
$\calH(x)=\ln x$ (eq.~(\ref{densityofstates}) above), this is exactly the same competition between
an exponentially growing density of states and an exponentially decaying
Boltzmann factor $e^{-\beta E}$ that characterizes the partition function of a
Hagedorn system (see Appendix B for details). The critical inverse temperature
(a.k.a.\ the Hagedorn inverse temperature),
$\beta_{\mbox{\tiny c}}=1$, is of course the boundary point between convergence and
divergence of the zeta function in the first place. 
It is well known that when $\beta$ is just slightly above 1, 
\begin{equation}
\label{criticalbehavior}
\zeta(\beta)\approx\frac{1}{\beta-1}
\end{equation}
in the sense that $\lim_{\beta\downarrow 1}(\beta-1)\zeta(\beta)=1$.

It should be noted that essentially the same effect occurs in physical models where the Hamiltonian
consists of a potential
function that is logarithmic in the position of the particle, similarly as in
\cite{KB10}, \cite{DLBK11}, \cite{HMS11}, \cite{HMS12}. However, here the
logarithmic Hamiltonian was obtained from a fundamental information-theoretic consideration, as
explained above. In an unrelated setting, the same qualitative link between
Zipf-law statistics and Hagedorn-type linear entropy-energy scaling (discussed
further in Sec.~1) has also been documented empirically in human fMRI brain
activity \cite{ChialvoJanik24}.

\subsubsection{The Bose Gas with Log-Prime Energy Levels}

We now recall a further statistical-mechanical analogy, which is well
established in the physics literature on arithmetic statistical mechanics but
is, to our knowledge, not commonly known within the information theory
community, and which we therefore include here for completeness and as a
further bridge between the two fields. The representation of $\zeta(\beta)$ as
the partition function of non-interacting bosons with energy levels equal to
logarithms of primes is precisely the {\em primon gas} (also called the {\em
Riemann gas}), introduced independently by Julia \cite{Julia90} and Spector
\cite{Spector90}, with the associated Hagedorn-type singularities studied
further by Julia \cite{Julia94}, and by Bakas and Bowick \cite{BakasBowick91};
see the colloquium review by Schumayer and Hutchinson \cite{SH11} for a
comprehensive account, including generalizations (Beurling gases,
supersymmetric and fermionic variants) that lie outside the present coding
context. What appears to be new in our treatment is not the arithmetic-gas
construction itself, but its explicit identification with the code-length
Hamiltonian $\calH(x)=\ln x$ that arises from the coding problem for the
integers, and the resulting connection between this established
statistical-mechanical picture and the zeta-distribution mismatch and
large-deviations analysis of Sec.~4, as well as the universal-coding
redundancy analysis of Sec.~5. Once the zeta
function is displayed in Euler's product form,
\begin{equation}
\zeta(\beta)=\prod_{p\in\calP}(1-p^{-\beta})^{-1}
=\prod_{p\in\calP}(1-e^{-\beta\ln p})^{-1},
\end{equation}
where $\calP=\{2,3,5,\ldots\}$ is the set of all primes.
The second expression is readily recognized as the grand-canonical partition function of
bosons with energy levels $E_p=\ln p$, $p\in\calP$, namely, again logarithmic
energy levels, except that here, not the logarithms of all positive integers are
involved, but only those of the primes. 
The mean occupation of state $p$ is therefore,
\begin{equation}
\bar{N}_p=\frac{1}{e^{\beta\ln p}-1},
\end{equation}
and so, the expected total number of particles is
\begin{equation}
\bar{N}=\sum_{p\in\calP}\frac{1}{e^{\beta\ln p}-1},
\end{equation}
which tends to infinity as $\beta\downarrow 1$ (by monotone convergence, each term increasing to $\frac{1}{p-1}$), since
$\sum_{p\in\calP}\frac{1}{p-1}\ge
\sum_{p\in\calP}\frac{1}{p}=\infty$, as was proved by Euler in 1737.
It therefore appears that from the viewpoint of the Bose gas model, the critical behavior
is manifested as a phenomenon of an unbounded increase in the number of
particles.

To understand the relation between the two statistical-mechanical models,
consider the factorization of the positive integer $x$ of the first model into its
prime factors,
\begin{equation}
x=\prod_{p\in\calP} p^{N_p},
\end{equation}
where $\{N_p\}_{p\in\calP}$ are non-negative integers, which designate the
multiplicities of the various prime factors. Now,
\begin{eqnarray}
\zeta(\beta)&=&\sum_{x=1}^\infty x^{-\beta}\nonumber\\
&=&\sum_{x=1}^\infty\prod_{p\in\calP} p^{-\beta N_p} \nonumber\\
&=&\sum_{N_2=0}^\infty\sum_{N_3=0}^\infty\sum_{N_5=0}^\infty\cdots\prod_{p\in\calP}p^{-\beta N_p}\nonumber\\
&=&\prod_{p\in\calP}\left(\sum_{k=0}^\infty p^{-\beta k}\right)\nonumber\\
&=&\prod_{p\in\calP}\left(1-p^{-\beta}\right)^{-1},
\end{eqnarray}
where the third equality is due to the fact every positive integer $x$ has
a unique factorization into prime factors. Consider an infinite series of
independent geometric random variables, $N_2, N_3, N_5,\ldots$, whose marginal
distributions are
\begin{equation}
\mbox{Pr}\{N_p=k\}=(1-p^{-\beta})p^{-\beta k},~~~k=0,1,2,\ldots.
\end{equation}
Then the random variable $X$ (under $P_\beta$) can be represented as
$\prod_{p\in\calP} p^{N_p}$, where each $N_p$, $p\in\calP$, is an independent
geometric random variable with parameter $p^{-\beta}$ (only finitely many $N_p$ are nonzero almost surely, by the Borel--Cantelli lemma, since $\sum_p\mbox{Pr}\{N_p\ge1\}=\sum_pp^{-\beta}<\infty$).
Accordingly, the occupation numbers, $\{N_p\}$, of this boson-gas model are simply the
multiplicities of the various primes in the factorization of $x$. Likewise, in
nats (matching the energy levels $E_p=\ln p$),
\begin{equation}
\ln X =\sum_{p\in\calP} N_p\ln p=\sum_{p\in\calP}N_pE_p.
\end{equation}
Converting to bits ($\log_2X=\ln X/\ln2=\sum_pN_p\log_2p$) to match the code
length $\ell(X)$, which is measured in bits throughout (Sec.~2),
\begin{equation}
\ell(X)=\beta\log_2X+\log_2\zeta(\beta)=\beta\sum_{p\in\calP} N_p\log_2p
-\sum_{p\in\calP}\log_2(1-p^{-\beta}),
\end{equation}
using $\log_2\zeta(\beta)=-\sum_p\log_2(1-p^{-\beta})$ from the Euler product.

\subsection{Multiple Particles}

\begin{figure}[h!t!b!]
\centering
\includegraphics[width=8.5cm, height=8.5cm]{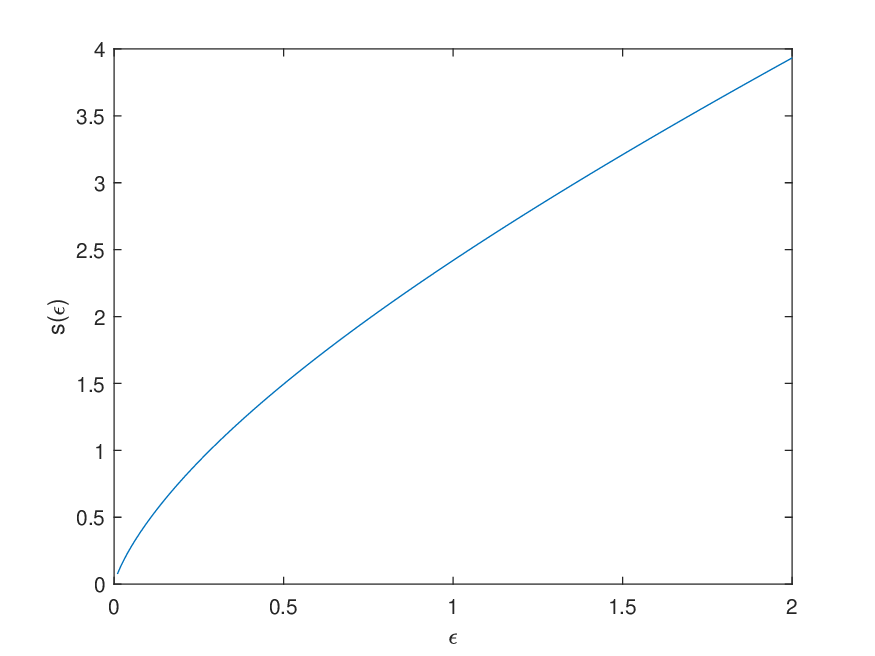}
\caption{The entropy function
$s(\epsilon)=\inf_{\beta>1}\{\beta\epsilon+\ln\zeta(\beta)\}$.
Observe that for large $\epsilon$, the function becomes nearly linear in
$\epsilon$.}
\label{graph1}
\end{figure}

We now return to the zeta function in its original form. 
For a single particle, we were able to observe an exponential density of
states only at very high energy levels.
To clarify the picture for all energy levels,
we now examine the density of
states, or more precisely, the micro-canonical entropy, in the case of
$N$ independent particles in the thermodynamic limit, $N\to\infty$.
Clearly,
\begin{equation}
[\zeta(\beta)]^N=\left[\sum_{x=1}^\infty e^{-\beta\ln x}\right]^N=
\sum_{x_1=1}^\infty 
\sum_{x_2=1}^\infty \ldots
\sum_{x_N=1}^\infty 
\exp\left\{-\beta\sum_{i=1}^N\ln x_i\right\}.
\end{equation}
Denoting the micro-state $(x_1,\ldots,x_N)$ by $\bx$, the Hamiltonian is now
given by
\begin{equation}
\calH(\bx)=\sum_{i=1}^N\ln x_i.
\end{equation}
The specific entropy, $s(\epsilon)$, for a given energy per particle,
$\epsilon$, is given by the Fenchel-Legendre transform of $\ln\zeta(\beta)$:
\begin{equation}
\label{entropydef}
s(\epsilon)=\inf_{\beta\ge 0}\{\beta\epsilon+\ln\zeta(\beta)\}
=\inf_{\beta>1}\{\beta\epsilon+\ln\zeta(\beta)\},
\end{equation}
where the second equality is due to the fact that $\zeta(\beta)$ diverges for
$\beta\in[0,1]$, and so, the infimum must be achieved (or at least approached)
in the range $\beta>1$. Fig. \ref{graph1} depicts the function $s(\epsilon)$
across a certain range of $\epsilon$, including small $\epsilon$. As can be
seen, for large $\epsilon$, the
curve becomes nearly linear. Indeed, for large $\epsilon$, the minimizing
$\beta$ approaches 1 (from above), and then
$\zeta(\beta)\approx\frac{1}{\beta-1}$. Therefore, for large $\epsilon$,
\begin{equation}
s(\epsilon)\approx\inf_{\beta>1}\{\beta\epsilon-\ln(\beta-1)\}.
\end{equation}
Taking the derivative of the right-hand side with respect to $\beta$ and
equating to zero, one readily finds that the minimizing $\beta$ is about
\begin{equation}
\label{betastar}
\beta^*=1+\frac{1}{\epsilon},
\end{equation}
which upon substituting back into the expression
$\beta^*\epsilon-\ln(\beta^*-1)$ yields
\begin{equation}
s(\epsilon)\approx \epsilon+\ln(e\cdot\epsilon),
\end{equation}
and so, the leading term is indeed linear in $\epsilon$ with coefficient
$\beta_{\mbox{\tiny c}}=1$, as expected. The corresponding density of states
for large $\epsilon$, is therefore, to leading exponential order,
$e^{Ns(\epsilon)}\approx(e\epsilon)^N e^{N\epsilon}$, and the Hagedorn-type behavior is observed once
again.

It is instructive to examine the equivalence between the micro-canonical and
canonical descriptions of this model precisely, following the standard
formalism (see, e.g., \cite{Touchette15}). The canonical mean energy per
particle, as a function of $\beta$ (for $\beta>1$ the series $\sum_xx^{-\beta}$ and its termwise derivatives converge absolutely and locally uniformly, so $\ln\zeta$ is real-analytic and may be differentiated termwise), is
\begin{equation}
\epsilon(\beta)=\bE_\beta[\ln X]=-\frac{\mbox{d}}{\mbox{d}\beta}\ln\zeta(\beta),
\qquad \beta>1,
\end{equation}
and differentiating once more gives
\begin{equation}
\frac{\mbox{d}\epsilon}{\mbox{d}\beta}=-\frac{\mbox{d}^2}{\mbox{d}\beta^2}\ln\zeta(\beta)
=-\mbox{Var}_\beta(\ln X)<0
\end{equation}
for every finite $\beta>1$. Hence $\epsilon(\beta)$ is continuous and strictly
decreasing on the entire canonical domain $\beta\in(1,\infty)$; since,
as $\beta\downarrow 1$, $\epsilon(\beta)\to\infty$ (otherwise the decreasing function $\epsilon$ would have a finite limit, and $\ln\zeta(\beta)=\ln\zeta(2)+\int_\beta^2\epsilon(t)\,\mbox{d}t$ would stay bounded, contradicting $\zeta(\beta)\to\infty$)
and $\epsilon(\beta)\to 0$ as $\beta\to\infty$ (where $P_\beta$ concentrates on
$x=1$), the map $\beta\mapsto\epsilon(\beta)$ is in fact a bijection from
$(1,\infty)$ onto $(0,\infty)$: every finite positive energy per particle
corresponds to a unique canonical inverse temperature. Correspondingly, for
every $\epsilon>0$ the function $\beta\mapsto\beta\epsilon+\ln\zeta(\beta)$ is
strictly convex (its second derivative is $\mbox{Var}_\beta(\ln X)>0$) and
diverges both as $\beta\downarrow 1$ and as $\beta\to\infty$, so the infimum
defining $s(\epsilon)$ in eq.\ (\ref{entropydef}) is attained at a unique
interior $\beta^*(\epsilon)$ satisfying $\epsilon=\epsilon(\beta^*(\epsilon))$.
By the envelope theorem, $s'(\epsilon)=\beta^*(\epsilon)$, and
\begin{equation}
s''(\epsilon)=\frac{\mbox{d}\beta^*}{\mbox{d}\epsilon}=
-\frac{1}{\mbox{Var}_{\beta^*}(\ln X)}<0,
\end{equation}
so $s(\epsilon)$ is strictly concave on $(0,\infty)$. Strict
concavity of the micro-canonical entropy together with the one-to-one energy
map are precisely the standard conditions for full thermodynamic equivalence
between the two ensembles \cite{Touchette15}; there is no plateau or affine
segment in $s(\epsilon)$ at any finite energy, and hence no failure of
equivalence on the finite-energy domain.

What {\em is} special about this model, and what motivates comparing it with a
Hagedorn system, is the behavior at the edge of that domain: as
$\epsilon\to\infty$, the corresponding inverse temperature $\beta^*(\epsilon)$
approaches $\beta_{\mbox{\tiny c}}=1$ but never reaches it at any finite
energy, consistent with the asymptotic relation $\beta^*(\epsilon)\approx
1+1/\epsilon$ derived above from eq.\ (\ref{betastar}). Equivalently, the
micro-canonical temperature $T_{\mbox{\tiny mc}}(\epsilon)=1/\beta^*(\epsilon)$
increases continuously toward the limiting value $T_{\mbox{\tiny c}}=1$ as
$\epsilon\to\infty$, but is strictly below it for every finite $\epsilon$; the
canonical ensemble is likewise restricted to $\beta>1$, i.e., $T<1$, so both
ensembles share exactly the same accessible temperature range. The appropriate
statement is therefore that $\beta_{\mbox{\tiny c}}=1$ (equivalently,
$T_{\mbox{\tiny c}}=1$) is a limiting temperature that can be approached but
never attained, exactly as in a Hagedorn system, and not a manifestation of
ensemble inequivalence.

For completeness, we verify that this conclusion is not an artifact of
identifying the micro-canonical entropy with the Legendre transform by
definition: the same conclusion follows from an independently defined
counting entropy, which we show below to coincide with $s(\epsilon)$. Define
\begin{equation}
\Omega_N(\epsilon,\delta)=\left|\left\{(x_1,\ldots,x_N)\in\calN^N:
\left|\frac1N\sum_{i=1}^N\ln x_i-\epsilon\right|<\delta\right\}\right|,
\end{equation}
\begin{eqnarray}
s^{+}_{\mbox{\tiny mc}}(\epsilon)&=&\lim_{\delta\downarrow0}\limsup_{N\to\infty}
\frac1N\ln\Omega_N(\epsilon,\delta),\\
s^{-}_{\mbox{\tiny mc}}(\epsilon)&=&\lim_{\delta\downarrow0}\liminf_{N\to\infty}
\frac1N\ln\Omega_N(\epsilon,\delta),
\end{eqnarray}
both limits in $\delta$ existing because $\Omega_N(\epsilon,\delta)$ is nondecreasing in $\delta$. We show below that, for every $\delta>0$, the $\limsup$ and the $\liminf$ over $N$ of $\frac1N\ln\Omega_N(\epsilon,\delta)$ lie between $s(\epsilon)-\beta^*(\epsilon)\delta$ and $s(\epsilon+\delta)$, so that $s^{+}_{\mbox{\tiny mc}}(\epsilon)\le s(\epsilon)\le s^{-}_{\mbox{\tiny mc}}(\epsilon)$ by continuity of $s$; since $s^{-}_{\mbox{\tiny mc}}\le s^{+}_{\mbox{\tiny mc}}$ trivially, the two coincide, and we denote their common value by $s_{\mbox{\tiny mc}}(\epsilon)$. For every $\beta>1$ we have the upper bound
\begin{equation}
\Omega_N(\epsilon,\delta)\le\zeta(\beta)^N e^{\beta N(\epsilon+\delta)},
\end{equation}
since each of the $\Omega_N(\epsilon,\delta)$ configurations in the shell
contributes at least $\zeta(\beta)^{-N}e^{-\beta N(\epsilon+\delta)}$ to
$P_\beta^{\otimes N}(\mbox{shell})\le1$. Taking $\beta>1$ to infimize the
right-hand side gives $\limsup_{N}\frac1N\ln\Omega_N(\epsilon,\delta)\le s(\epsilon+\delta)$, and hence, letting $\delta\downarrow0$ and using the continuity of $s$, $s^{+}_{\mbox{\tiny mc}}(\epsilon)\le s(\epsilon)$. For the
matching lower bound, take $\beta=\beta^*(\epsilon)$, the unique minimizer
identified above. By the (weak) law of large numbers under
$P_{\beta^*}^{\otimes N}$, since $\bE_{\beta^*}[\ln X]=\epsilon$, the
shell $\{|\frac1N\sum\ln X_i-\epsilon|<\delta\}$ has
$P_{\beta^*}^{\otimes N}$-probability tending to $1$ as $N\to\infty$, for any
fixed $\delta>0$; since every configuration in the shell has
$P_{\beta^*}^{\otimes N}$-probability at most
$\zeta(\beta^*)^{-N}e^{-\beta^*N(\epsilon-\delta)}$, this gives
$\Omega_N(\epsilon,\delta)\ge(1-o(1))\,\zeta(\beta^*)^Ne^{\beta^*N(\epsilon-\delta)}$,
so $\liminf_N\frac1N\ln\Omega_N(\epsilon,\delta)\ge\beta^*(\epsilon)\epsilon+\ln\zeta(\beta^*(\epsilon))-\beta^*(\epsilon)\delta=s(\epsilon)-\beta^*(\epsilon)\delta$
for every $\delta>0$; letting $\delta\downarrow0$ gives
$s^{-}_{\mbox{\tiny mc}}(\epsilon)\ge s(\epsilon)$. Combining the two bounds,
$s_{\mbox{\tiny mc}}(\epsilon)=s(\epsilon)$: the independently defined
counting entropy coincides with the Legendre transform, and the equivalence
and limiting-temperature conclusions above apply equally to it.

By contrast, if the particle states are limited to a finite range,
$\{1,2,\ldots,J\}$, the partition function $\sum_{x=1}^J x^{-\beta}$ converges
for every real $\beta$, including $\beta\le 0$; allowing the full range
$\beta\in\reals$, the curve of $s(\epsilon)$ rises from $0$ at $\epsilon=0$ to
its maximum $\ln J$ at the uniform-distribution energy and then decreases,
recovering a one-to-one map between $\beta$ and $\epsilon$ over the whole
finite energy range, with no limiting temperature.\footnote{If the Legendre
transform is instead restricted to $\beta\ge 0$, as one might do by analogy
with the infinite-support case, one obtains a spurious plateau
$s(\epsilon)=\ln J$ for $\epsilon$ above the uniform-distribution energy, since
the constrained minimizer then sits at the boundary $\beta=0$ for all such
$\epsilon$; this plateau is an artefact of the unnecessary restriction
$\beta\ge 0$ and disappears once negative $\beta$ is allowed.}

\subsubsection{A Sub-Leading Correction and a Critical Exponent}

The leading-order relations above can be sharpened using one further classical
fact about $\zeta(\beta)$: its Laurent expansion at the simple pole
$\beta=1$,
\begin{equation}
\label{laurent}
\zeta(\beta)=\frac{1}{\beta-1}+\gamma+O(\beta-1),\qquad\beta\to 1^+,
\end{equation}
where $\gamma=0.5772\ldots$ is the Euler--Mascheroni constant
\cite{Apostol76}. Since $\zeta(s)$ is meromorphic on all of $\mathbb{C}$ with its
\emph{only} singularity being the simple pole at $s=1$, the function
$g(s):=\zeta(s)-\frac{1}{s-1}$ is entire, with $g(1)=\gamma$, and
$\zeta(s)=(s-1)^{-1}\bigl[1+(s-1)g(s)\bigr]$. Since $1+(s-1)g(s)\to1\ne0$ as $s\to1$,
$h(s):=\ln\bigl[1+(s-1)g(s)\bigr]$ is analytic in a neighborhood of $s=1$, with
$h(1)=0$ and $h'(1)=g(1)=\gamma$; its Taylor series converges there and may be
differentiated term by term. Hence $\ln\zeta(\beta)=-\ln(\beta-1)+h(\beta)$ and, as $\beta\to 1^+$,
\begin{equation}
\ln\zeta(\beta)=-\ln(\beta-1)+\gamma(\beta-1)+O((\beta-1)^2),
\end{equation}
with $h'$ and $h''$ having the corresponding expansions (so that the error terms below may be obtained by differentiating the expansion),
and hence, recalling $\epsilon(\beta)=-\mbox{d}\ln\zeta(\beta)/\mbox{d}\beta$
and $\mbox{Var}_\beta(\ln X)=\mbox{d}^2\ln\zeta(\beta)/\mbox{d}\beta^2$
from above,
\begin{eqnarray}
\label{epsrefined}
\epsilon(\beta)&=&\frac{1}{\beta-1}-\gamma+O(\beta-1),\\
\label{varexponent}
\mbox{Var}_\beta(\ln X)&=&\frac{1}{(\beta-1)^2}+O(1),\qquad\beta\to 1^+.
\end{eqnarray}
Two consequences follow. First, (\ref{varexponent}) shows that the energy
fluctuations per particle -- the direct analog of specific heat in this
model -- diverge at the Hagedorn temperature precisely as
$(\beta-\beta_{\mbox{\tiny c}})^{-2}$. This exponent is not particular to the
zeta distribution: it follows purely from $\zeta(\beta)$ having a
\emph{simple} (order-one) pole at $\beta_{\mbox{\tiny c}}=1$, so any
system whose partition function is meromorphic near its critical inverse temperature
with a simple pole there (more generally, $\ln Z=-\ln(\beta-\beta_{\mbox{\tiny c}})+$ a $C^2$ remainder) will exhibit the same
critical exponent $2$ for this quantity, regardless of the microscopic
details producing that pole.

Second, inverting $\epsilon(\beta^*(\epsilon))=\epsilon$ to next order using
(\ref{epsrefined}) gives the refined estimate
\begin{equation}
\beta^*(\epsilon)=1+\frac{1}{\epsilon+\gamma}+O(1/\epsilon^2),
\end{equation}
sharpening $\beta^*(\epsilon)\approx 1+1/\epsilon$ used above, and
substituting back into $s(\epsilon)=\beta^*(\epsilon)\epsilon+\ln\zeta(\beta^*(\epsilon))$
yields the refined micro-canonical entropy asymptotic
\begin{equation}
\label{entropyrefined}
s(\epsilon)=\epsilon+1+\ln(\epsilon+\gamma)+O(1/\epsilon^2),\qquad\epsilon\to\infty,
\end{equation}
sharpening the leading-order estimate
$s(\epsilon)\approx\epsilon+\ln(e\epsilon)=\epsilon+1+\ln\epsilon$ obtained
above. Table~\ref{tab:entropy} compares (\ref{entropyrefined}) against the
entropy $s(\epsilon)$ computed by direct numerical Legendre transform (solving
$\epsilon(\beta^*)=\epsilon$ for $\beta^*$ to $30$ significant digits and
evaluating $\beta^*\epsilon+\ln\zeta(\beta^*)$): including the
Euler--Mascheroni correction reduces the approximation error by a factor of about $36$ at $\epsilon=5$, $67$ at
$\epsilon=10$ and over $300$ at $\epsilon=50$.

\begin{table}[h!t!b!]
\centering
\begin{tabular}{c c c c}
\hline
$\epsilon$ & $s(\epsilon)$ (exact) & $\epsilon+1+\ln\epsilon$ & $\epsilon+1+\ln(\epsilon+\gamma)$ \\
\hline
5   & 7.715754  & 7.609438 \ (err $1.1\times10^{-1}$) & 7.718690 \ (err $3\times10^{-3}$) \\
10  & 13.357877 & 13.302585 (err $5.5\times10^{-2}$) & 13.358702 (err $8\times10^{-4}$) \\
20  & 24.023965 & 23.995732 (err $2.8\times10^{-2}$) & 24.024184 (err $2\times10^{-4}$) \\
50  & 54.923465 & 54.912023 (err $1.1\times10^{-2}$) & 54.923501 (err $4\times10^{-5}$) \\
\hline
\end{tabular}
\caption{Numerical verification of the refined entropy asymptotic
(\ref{entropyrefined}) against the exact micro-canonical entropy, computed by
solving $\epsilon(\beta^*)=\epsilon$ numerically. The Euler--Mascheroni
correction reduces the error by a factor of about $36$ at $\epsilon=5$ and of over $300$ at $\epsilon=50$, relative to
the leading-order estimate.}
\label{tab:entropy}
\end{table}

\section{Large Deviations Analysis}

It is instructive to examine also the large deviations behavior associated
with the zeta distribution: What is the exponential rate of the probability
that $\sum_{i=1}^N\ln X_i$ would exceed $N\epsilon$, where $\epsilon>0$ is a
given constant, independent of $N$, and larger than the expectation of $\ln
X_1$. This is readily accomplished by applying
the Chernoff bound:
\begin{eqnarray}
& &\mbox{Pr}\left\{\sum_{i=1}^N\ln X_i\ge N\epsilon\right\}\nonumber\\
&\le&\inf_{\lambda\ge 0}\bE\left\{\exp\left[\lambda\left(\sum_{i=1}^N\ln
X_i-N\epsilon\right)\right]\right\}\nonumber\\
&=&\inf_{\lambda\ge 0} e^{-\lambda\epsilon
N}\bE\left\{\exp\left[\lambda\sum_{i=1}^N\ln X_i\right]\right\}\nonumber\\
&=&\inf_{\lambda\ge 0} e^{-\lambda\epsilon
N}\left[\bE\left\{\exp\left(\lambda\ln X\right)\right\}\right]^N\nonumber\\
&=&\inf_{\lambda\ge 0} e^{-\lambda\epsilon
N}\left[\bE\{X^\lambda\}\right]^N\nonumber\\
&=&\inf_{0\le\lambda<\beta-1} e^{-\lambda\epsilon
N}\left[\frac{\zeta(\beta-\lambda)}{\zeta(\beta)}\right]^N\nonumber\\
&=&\exp\left\{-N\cdot\sup_{0\le\lambda<\beta-1}\left[\lambda\epsilon-\ln\zeta(\beta-\lambda)+\ln\zeta(\beta)\right]\right\}.
\end{eqnarray}
The derivation above is, strictly speaking, only a Chernoff upper bound on the
large deviations probability. It is, however, exponentially tight: the
logarithmic moment generating function $\Lambda_\beta(\lambda)=\ln
\zeta(\beta-\lambda)-\ln\zeta(\beta)$ of $\ln X$ under $P_\beta$ is finite on
the open interval $\lambda\in(-\infty,\beta-1)$ containing the origin,
so Cram\'er's theorem applies directly
and the exponent above,
$I_\beta(\epsilon)=\sup_{\lambda\ge 0}\{\lambda\epsilon-\Lambda_\beta(\lambda)\}$,
is the exact large-deviations rate for every $\epsilon$ larger than the mean
$\mu_\beta=\bE_\beta[\ln X]$; for $\epsilon<\mu_\beta$ the event is typical
(probability tending to 1) and for $\epsilon=\mu_\beta$ the rate is zero.

This rate function admits a compact restatement directly in terms of the
micro-canonical entropy $s(\epsilon)$ of Sec.~3. Writing $\tau^*(\epsilon)$
for the minimizer in the Legendre transform
$s(\epsilon)=\inf_{\nu>1}\{\nu\epsilon+\ln\zeta(\nu)\}$ (so that
$\bE_{\tau^*(\epsilon)}[\ln X]=\epsilon$; $\tau^*(\epsilon)$ is the function denoted $\beta^*(\epsilon)$ in Sec.~3), a direct calculation using the
definition of $I_\beta(\epsilon)$ above gives, for every $\epsilon\ge\mu_\beta$, the exact identity
\begin{equation}
\label{cramerKL}
I_\beta(\epsilon)=\beta\epsilon+\ln\zeta(\beta)-s(\epsilon)
=D\bigl(P_{\tau^*(\epsilon)}\,\|\,P_\beta\bigr),
\end{equation}
i.e., for $\epsilon\ge\mu_\beta$ (the range in which the event is atypical, and in which $\tau^*(\epsilon)\le\beta$; for $\epsilon<\mu_\beta$ the rate is zero) the Cram\'er rate is exactly the Kullback--Leibler divergence of the
exponentially tilted distribution $P_{\tau^*(\epsilon)}$, under which the atypical value $\epsilon$ becomes
typical, from the true source $P_\beta$. The large-deviations exponent of this section is thus, in an exact
sense, nothing but the Legendre-transform entropy of Sec.~3, viewed from the
perspective of the (untilted) true source.

We note that, for any fixed $\epsilon$, the large deviations rate function
vanishes whenever the mean $\mu_\beta=\bE_\beta[\ln X]$ reaches $\epsilon$ (the event is then typical), which happens for all $\beta\le\tau^*(\epsilon)$, since $\mu_\beta\to\infty$ as $\beta\downarrow1$; correspondingly the range of optimization of the Chernoff
parameter $\lambda$ shrinks to zero as $\beta\downarrow 1$. When $\epsilon$ is large,
the optimal $\lambda$
is about $\lambda^*\approx\beta-1-\frac{1}{\epsilon}$, and the resulting large deviations rate
function becomes nearly
$\epsilon(\beta-1)-\ln(e\cdot\epsilon)+\ln\zeta(\beta)$, namely, essentially
linear in $\epsilon$ for large $\epsilon$. Correspondingly, the tilted
parameter $\tau^*(\epsilon)=\beta-\lambda^*$ satisfies $\tau^*(\epsilon)-1\sim 1/\epsilon$, so its
approach to the critical value $1$ as $\epsilon\to\infty$ is algebraic
(power-law), not exponential, in $\epsilon$; we return to this point in the
Summary.

\begin{figure}[h!t!b!]
\centering
\includegraphics[width=8.5cm, height=8.5cm]{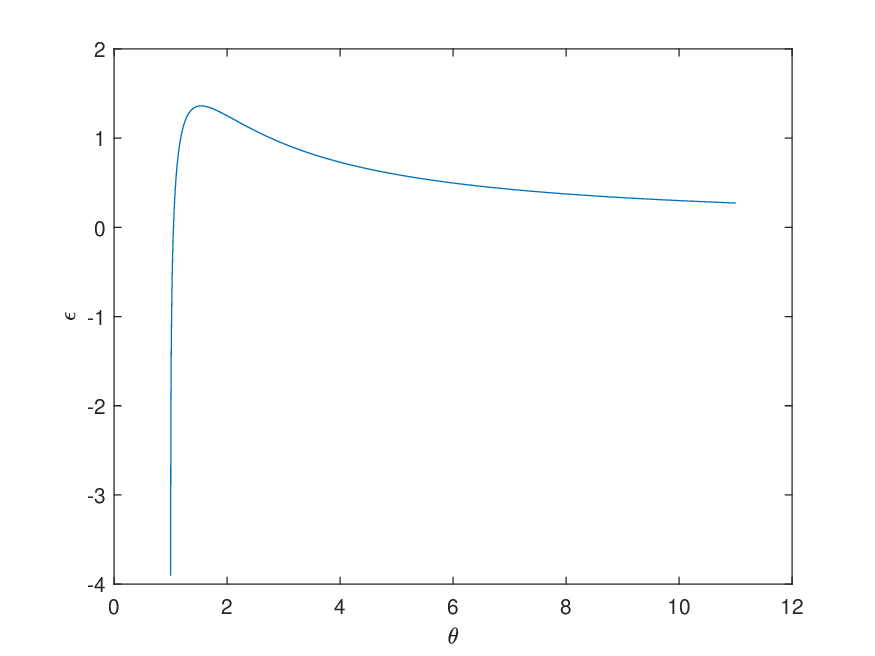}
\caption{A graph of $\epsilon$ vs.\ $\theta$ for $R\ln 2=3$.}
\label{graph2}
\end{figure}

The large deviations behavior is relevant also for the coding problem
discussed earlier, and is closely related to the classical analysis of buffer
overflow in variable-length coding of fixed-rate sources \cite{Jelinek68}.
Suppose that we encode a block $\bX=(X_1,\ldots,X_N)$ of
independent integers, all governed by $P_\beta$. What is the probability
that the total code-length $\ell(\bX)=-\log
P_\beta(\bX)=-\sum_{i=1}^N\log P_\beta(X_i)$ (in bits) exceeds a certain
threshold, $NR$. (As in Sec.~2, integer-length constraints are neglected here; coding the whole block with ceiling lengths adds less than one bit to the total, which does not affect the exponent.) The motivation is clear: suppose we wish to store the
compressed representation of $\bX$ in a buffer of size $NR$ bits and we are
concerned by the unfortunate event of buffer overflow, which causes loss of information.
Based on the structure of $P_\beta$, this is the event $\beta\sum_{i=1}^N\log_2
X_i+N\log_2\zeta(\beta)\ge NR$, or equivalently (converting to nats and
dividing through by $N$),
\begin{equation}
\frac{1}{N}\sum_{i=1}^N\ln X_i\ge\frac{R\ln 2-\ln\zeta(\beta)}{\beta}.
\end{equation}
But when it comes to large deviations performance, the length function
$\ell(\bX)=-\log P_\beta(\bX)$
may not be optimal in the sense of maximizing the large-deviations rate
function. It is instructive to examine whether there is another value of the
parameter of the zeta distribution, call it $\theta$, whose corresponding length function,
$\ell_\theta(\bX)=-\log P_{\theta}(\bX)$ would be better in that sense (we compare only within the family $\{\ell_\theta:\theta>1\}$).  
For $\theta$, the buffer overflow event becomes
$\theta\sum_{i=1}^N\log_2 X_i+N\log_2\zeta(\theta)\ge NR$, or, again converting
to nats and dividing through by $N$,
\begin{equation}
\frac{1}{N}\sum_{i=1}^N\ln X_i\ge\frac{R\ln 2-\ln\zeta(\theta)}{\theta},
\end{equation}
and then we are back to the above derivation with the assignment
\begin{equation}
\epsilon=\epsilon_R(\theta):= \frac{R\ln 2-\ln\zeta(\theta)}{\theta}.
\end{equation}
The optimal value of $\theta$, denoted $\theta^*$, is the one that maximizes the right-hand
side of the last equation, because the rate $I_\beta(\epsilon)$ is increasing in $\epsilon$ for $\epsilon>\mu_\beta$ (and the event is typical otherwise). Interestingly, this value of $\theta$ depends
merely on $R$, and not on the parameter $\beta$ of the underlying zeta
distribution. This means that when designing a code with good large
deviations behavior, one needs to know merely the buffer size, but not the
parameter $\beta$ of the underlying source (the resulting exponent $I_\beta(\epsilon_{\max})$ itself, of course, still depends on $\beta$).
Fig \ref{graph2} depicts the curve of $\epsilon$ as a function of
$\theta$ for $R\ln 2=3$. The maximum achieved is $\epsilon_{\max}=1.3617$
at $\theta=1.54$.

Differentiating $\epsilon_R(\theta)$ and setting the result to zero shows
that the maximizing $\theta$ satisfies the stationarity condition
\begin{equation}
\theta\,\bE_\theta[\ln X]+\ln\zeta(\theta)=R\ln2,
\qquad\mbox{i.e.}\qquad H(P_\theta)=R\ln2,
\end{equation}
($H(P_\theta)$ in nats), with the achieved value
$\epsilon_{\max}=\bE_\theta[\ln X]$ at that $\theta$ (confirmed by the
numbers above: for $R\ln2=3$, i.e., $R=3/\ln2\approx4.328$ bits, the root of $H(P_\theta)=3$ is
$\theta^*=1.540343\ldots$, at which $\bE_{\theta^*}[\ln X]=1.361705\ldots$, matching $\theta=1.54$ and
$\epsilon_{\max}=1.3617$ to the precision quoted). Since $H(P_\theta)=s\bigl(\bE_\theta[\ln X]\bigr)$, where $s$ is increasing ($s'(\epsilon)=\tau^*(\epsilon)>0$) and $\bE_\theta[\ln X]$ is decreasing in $\theta$ (its derivative is $-\mbox{Var}_\theta(\ln X)<0$),
$H(P_\theta)$ is strictly decreasing in $\theta$, from $+\infty$ as $\theta\downarrow1$ to $0$ as $\theta\to\infty$; hence, for every $R>0$, the equation $H(P_\theta)=R\ln2$ has a unique root and this stationary point is unique; it is the maximizer, since
$\epsilon_R(\theta)\to-\infty$ as $\theta\downarrow1$ and $\epsilon_R(\theta)\to0$ as $\theta\to\infty$. Moreover, $H(P_\beta)<R\ln2=H(P_{\theta^*})$ if and only if
$\theta^*<\beta$, i.e., if and only if $\epsilon_{\max}=\bE_{\theta^*}[\ln X]>\bE_\beta[\ln X]$, so the condition $R>H(P_\beta)/\ln2$ stated next
applies to the optimal $\theta^*$ as well as to the matched $\theta=\beta$. This also makes explicit the condition for a positive
overflow exponent in the matched case $\theta=\beta$: since $\epsilon_R(\beta)$
exceeds the true mean $\mu_\beta=\bE_\beta[\ln X]$ exactly when
$R\ln2>H(P_\beta)$,
the overflow event has a positive large-deviations exponent under the
matched source $P_\beta$ if and only if $R>H(P_\beta)/\ln2$ (bits); otherwise
the overflow event is typical and the exponent is zero -- cf.\ the
buffer-overflow criterion in \cite{Jelinek68,Merhav91overflow}, here recovered
directly from the present formalism.

\section{Universal Coding and the Jeffreys Integral at the Hagedorn Transition}

We close the technical development with a further, genuinely
information-theoretic consequence of the critical exponent derived in
Sec.~3.2.1, connecting back to the universal-coding motivation of Sec.~1.

The zeta distribution $P_\beta(x)=x^{-\beta}/\zeta(\beta)$ is a
one-dimensional exponential family with sufficient statistic $\ln x$,
natural parameter $\beta\in(1,\infty)$, and log-partition function
$\ln\zeta(\beta)$. Its Fisher information (for estimating $\beta$ from a
single observation) is therefore exactly the same quantity already computed
in Sec.~3.2.1:
\begin{equation}
F(\beta)=\mathrm{Var}_\beta(\ln X)=\frac{d^2}{d\beta^2}\ln\zeta(\beta)
=\frac{1}{(\beta-1)^2}+O(1),\qquad\beta\to1^+.
\end{equation}
Now suppose $\beta$ is unknown, and we wish to design a single universal
code for an i.i.d.\ $N$-tuple $X^N=(X_1,\ldots,X_N)$ drawn from $P_\beta$ --
the same sample-size notation $N$ as in the large-deviations analysis of
Sec.~4 -- that performs well simultaneously for every $\beta$ in some range.
For a one-dimensional exponential family with Fisher information
$F(\beta)$, restricted to a compact set $K=[a,b]$ in the interior of the
parameter space, the result of Clarke and Barron \cite{BarronClarke94} gives the asymptotic \emph{minimax expected redundancy}
$R^{\mathrm{red}}_N(K):=\inf_{Q_N}\sup_{\beta\in K}D\bigl(P_\beta^{\otimes N}\|Q_N\bigr)$ of
universal coding of $N$ i.i.d.\ observations as (in nats, i.e.\ using
$\ln$, as is standard for this classical result -- unlike the code lengths
$\ell(x)$ of Sec.~2, which are in bits)
\begin{equation}
\label{cbredundancy}
R^{\mathrm{red}}_N(K)=\frac12\ln\frac{N}{2\pi e}+\ln\int_a^b\sqrt{F(\beta)}\,d\beta+o(1),
\end{equation}
provided the Jeffreys integral $J(K):=\int_a^b\sqrt{F(\beta)}\,d\beta$ is finite.
The \emph{minimax regret} is
$R^{\mathrm{reg}}_N(K):=\inf_{Q_N}\sup_{x^N}\ln\bigl[\sup_{\beta\in K}P_\beta^{\otimes N}(x^N)/Q_N(x^N)\bigr]$,
and always $R^{\mathrm{red}}_N(K)\le R^{\mathrm{reg}}_N(K)$, since an expectation never exceeds a maximum.

Since $\sqrt{F(\beta)}\sim1/(\beta-1)$ as $\beta\to1^+$, this integral is
not finite once $a=1$ is included. It is tempting to conclude directly that
the minimax regret and redundancy are therefore infinite at the critical point, but this
step requires care: while divergence of the Shtarkov (NML) normalizing sum
always forces divergence of the Jeffreys integral, the converse
implication -- the one actually needed here -- fails in general for
one-sided exponential families; Gr\"unwald and Harremo\"es
\cite{GrunwaldHarremoes09} give an explicit counterexample (a family with
bounded redundancy but an infinite Jeffreys integral). We therefore
establish the conclusion directly for the zeta family, for both the minimax regret and the
minimax expected redundancy, and for every sample size $N$.

We begin with the single-observation ($N=1$) problem, first for the regret and then for the expected
redundancy; it turns out to already force the conclusion for every $N$. Here the relevant parameter
range is the full punctured interval $(1,\infty)$ itself (not yet a
compact margin set $[1+\delta,b]$ as below): any range bounded
away from $\beta=1$, say $K\subseteq[1+\delta,\infty)$, gives a finite $N$-sample Shtarkov sum, since
$\sup_{\beta\ge1+\delta}P_\beta^{\otimes N}(x^N)\le\prod_{i}x_i^{-(1+\delta)}$ (because $x^{-\beta}\le x^{-(1+\delta)}$ and $\zeta(\beta)\ge1$), whose sum over
$x^N$ is $\zeta(1+\delta)^N$, so it
is essential that $\beta$ be allowed to approach $1$ freely here. 

For
$x\in\calN$, let $M(x)=\sup_{\beta>1}P_\beta(x)$, so that $\sum_xM(x)$ is
exactly the one-sample Shtarkov normalizing sum for this range. By
Shtarkov's theorem \cite{Shtarkov87} this is not merely an implication but
an exact identity: writing $R^{\mathrm{reg}}_1$ for the
one-sample minimax regret $R^{\mathrm{reg}}_1\bigl((1,\infty)\bigr)$, the normalized-maximum-likelihood
code $Q^*(x):=M(x)/\sum_{x'}M(x')$ achieves
\begin{equation}
\label{shtarkovexact}
R^{\mathrm{reg}}_1=\ln\sum_xM(x)
\end{equation}
whenever the sum is finite, and $R^{\mathrm{reg}}_1=\infty$ precisely when it is not --
with no asymptotic approximation involved at any stage. 

For $x\ge2$ the maximizing
$\beta_x$ satisfies the stationarity condition
$\epsilon(\beta_x)=\ln x$ (the function $\beta\mapsto\ln P_\beta(x)=-\beta\ln x-\ln\zeta(\beta)$ is concave, since $\ln\zeta$ is convex, so for $x\ge2$ this is the unique maximizer; for $x=1$ the supremum $M(1)=1$ is only approached as $\beta\to\infty$), i.e.\ $\beta_x=\beta^*(\ln x)$ in the notation of
Sec.~3.2.1, so (\ref{epsrefined}) gives
$\beta_x-1=\frac{1}{\ln x+\gamma}+O(1/(\ln x)^2)$. Substituting into
$\ln M(x)=-\beta_x\ln x-\ln\zeta(\beta_x)$ and using the Laurent expansion
(\ref{laurent}) once more yields, after simplification,
\begin{equation}
M(x)=\frac{1}{ex(\ln x+\gamma)}\bigl(1+O(1/\ln x)\bigr),\qquad x\to\infty,
\end{equation}
confirmed numerically (the ratio to the exact supremum is within $0.5\%$
already at $x=50$). Since $\sum_x\frac{1}{x(\ln x+\gamma)}$ diverges -- it
grows like $\ln\ln x$, by the integral test -- the one-sample Shtarkov sum
diverges, and (\ref{shtarkovexact}) gives $R^{\mathrm{reg}}_1=\infty$. The identical
conclusion holds for $R^{\mathrm{reg}}_1(K)$ with $\beta$ restricted to \emph{any}
sub-interval $K\subseteq(1,\infty)$ with $\inf K=1$ (not merely
$K=(1,\infty)$ itself): since $\beta_x\to1^+$, such a $K$ contains
$\beta_x$ for all sufficiently large $x$, so $\sup_{\beta\in K}P_\beta(x)$
agrees with the unrestricted $M(x)$ for all large $x$ -- the two series
differ in only finitely many terms, hence diverge together. A compact
margin set $K=[1+\delta,b]$ with $\delta>0$, by contrast, is bounded away
from $1$ and gives a convergent Shtarkov sum, consistent with
(\ref{cbredundancy}) applying there.

{\em $N$ samples, regret.} Let $S_N(K):=\sum_{x^N}\sup_{\beta\in K}P_\beta^{\otimes N}(x^N)$ be the $N$-sample Shtarkov sum, so that
$R^{\mathrm{reg}}_N(K)=\ln S_N(K)$ by Shtarkov's theorem. Since a supremum of a sum is at most the sum of the suprema,
\begin{eqnarray}
S_N(K)&=&\sum_{x_1}\sum_{x_2,\ldots,x_N}\sup_{\beta\in K}P_\beta^{\otimes N}(x^N)\nonumber\\
&\ge&\sum_{x_1}\sup_{\beta\in K}\sum_{x_2,\ldots,x_N}P_\beta^{\otimes N}(x^N)\;=\;\sum_{x_1}\sup_{\beta\in K}P_\beta(x_1)\;=\;S_1(K),
\end{eqnarray}
where the middle equality holds because summing $P_\beta^{\otimes N}$ over $x_2,\ldots,x_N$ leaves the marginal $P_\beta(x_1)$.
Hence $S_1(K)=\infty$ implies $S_N(K)=\infty$ and $R^{\mathrm{reg}}_N(K)=\infty$ for every $N\ge1$.

{\em Expected redundancy, one sample.} The Shtarkov identity concerns the regret, which is at least the expected redundancy, so we give a separate short
argument for the latter.
Let $K\subseteq(1,\infty)$ with $\inf K=1$.

{\em Claim:} $R^{\mathrm{red}}_1(K)=\infty$. {\em Proof.} Fix $L\ge2$. Since $\inf K=1$, we can choose $\beta_j=1+\lambda_j\in K$, $j=0,\ldots,L-1$, with $\lambda_0\le\frac1{10}$ and $\lambda_{j+1}\le\lambda_j/40$. Let
$A_j=\{x:\,e^{0.1/\lambda_j}\le x<e^{4/\lambda_j}\}$, $j=0,\ldots,L-1$.
The $A_j$ are pairwise disjoint, since $e^{4/\lambda_j}\le e^{0.1/\lambda_{j+1}}$, and $P_{\beta_j}(A_j)\ge0.62$: indeed, with $\beta=1+\lambda$,
$1/\lambda<\zeta(\beta)<1+1/\lambda$ (compare $\sum_{x\ge1}x^{-\beta}$ and $\sum_{x\ge2}x^{-\beta}$ with $\int_1^\infty u^{-\beta}du=1/\lambda$) and comparison
of the tails with integrals give, for $a:=e^{0.1/\lambda}\ge e$ and $n:=\lceil a\rceil\le a+1$,
$\sum_{x\ge n}x^{-\beta}\ge n^{-\lambda}/\lambda\ge a^{-\lambda}e^{-\lambda}/\lambda$, hence $P_\beta(X\ge a)\ge e^{-0.1-\lambda}/(1+\lambda)\ge0.74$; and, for $u:=e^{4/\lambda}$ and $n:=\lceil u\rceil\ge u$,
$\sum_{x\ge n}x^{-\beta}\le n^{-\beta}+n^{-\lambda}/\lambda$, hence $P_\beta(X\ge u)\le\lambda+e^{-4}<0.12$. Now let $Q$ be any one-sample code. Since the $A_j$ are disjoint, $\sum_jQ(A_j)\le1$, so $Q(A_j)\le1/L$ for at least one $j$; fix such a $j$.
Write $p=P_{\beta_j}(A_j)\ge\frac12$ and $q=Q(A_j)\le\frac1L$. The divergence cannot increase when $x$ is replaced by the
indicator $\mathbf 1\{x\in A_j\}$ (data processing), so
\begin{eqnarray}
D(P_{\beta_j}\|Q)&\ge&p\ln\frac pq+(1-p)\ln\frac{1-p}{1-q}\nonumber\\
&=&\bigl[p\ln p+(1-p)\ln(1-p)\bigr]+p\ln\frac1q+(1-p)\ln\frac1{1-q}\nonumber\\
&\ge&-\ln2+\frac12\ln L+0 .
\end{eqnarray}
The last step uses three facts: the bracket is minus the binary entropy of $p$, hence $\ge-\ln2$; $p\ln\frac1q\ge\frac12\ln L$ because
$p\ge\frac12$ and $q\le\frac1L$; and $(1-p)\ln\frac1{1-q}\ge0$.
Hence $\sup_{\beta\in K}D(P_\beta\|Q)\ge\frac12\ln L-\ln2$ for every $Q$ and every $L$, i.e., $R^{\mathrm{red}}_1(K)=\infty$. $\Box$

{\em Expected redundancy, $N$ samples.} The $N$-sample problem follows from the one-sample claim by a short monotonicity argument. For any candidate code (probability
distribution) $Q_N$ on $x^N$, let $Q_1$ be its first-coordinate marginal,
$Q_1(x_1):=\sum_{x_2,\ldots,x_N}Q_N(x^N)$. Since $P_\beta^{\otimes N}$'s
first-coordinate marginal is $P_\beta$ itself (the observations are
i.i.d.), the data-processing inequality for the Kullback--Leibler
divergence under the (deterministic) map $x^N\mapsto x_1$ \cite{CT06}
gives, for every $\beta$,
\begin{equation}
D\bigl(P_\beta^{\otimes N}\,\|\,Q_N\bigr)\ge D\bigl(P_\beta\,\|\,Q_1\bigr).
\end{equation}
Taking $\sup_{\beta\in K}$ of both sides, and noting that $Q_1$, being some valid one-sample code, satisfies
$\sup_{\beta\in K}D(P_\beta\|Q_1)\ge R^{\mathrm{red}}_1(K)$ by definition of the latter as an
infimum over \emph{all} one-sample codes, we obtain $\sup_{\beta\in K}D(P_\beta^{\otimes N}\|Q_N)\ge R^{\mathrm{red}}_1(K)$
for \emph{every} $N$-sample code $Q_N$. Taking $\inf_{Q_N}$ of the left-hand side yields
\begin{equation}
\label{regretmonotone}
R^{\mathrm{red}}_N(K)=\inf_{Q_N}\sup_{\beta\in K}D\bigl(P_\beta^{\otimes N}\|Q_N\bigr)
\ge R^{\mathrm{red}}_1(K),
\end{equation}
for every $N\ge1$ and every parameter range $K$. Combined with $R^{\mathrm{red}}_1(K)=\infty$ above,
(\ref{regretmonotone}) gives $R^{\mathrm{red}}_N(K)=\infty$ for every $N\ge1$, for every
$K\subseteq(1,\infty)$ with $\inf K=1$. Universal coding of an i.i.d.\
$N$-tuple, for any sample size, and the physical model become singular at
exactly the same point, for exactly the same reason -- the divergence of
$\mathrm{Var}_\beta(\ln X)$.

We apply (\ref{cbredundancy}) over the compact range $\beta\in[1+\delta,30]$, which lies within its hypothesis (the choice of a finite upper limit affects only the constant $C$ below; no claim is made for an unbounded range). Write $J(\delta):=\int_{1+\delta}^{30}\sqrt{F(\beta)}\,d\beta$
for this specific Jeffreys integral.

We now establish the asymptotic form of $J(\delta)$ rigorously, rather than
leaving it as an empirical observation. Write $u=\beta-1$. Since
$u\zeta(1+u)=1+\gamma u+c_2u^2+c_3u^3+\cdots$ is a convergent power series in
a neighborhood of $u=0$ -- the same fact used in deriving (\ref{laurent}) --
its logarithm is too:
\begin{equation}
\ln\bigl[u\,\zeta(1+u)\bigr]=\gamma u+d_2u^2+d_3u^3+\cdots=:\gamma u+u^2\rho(u),
\end{equation}
where $\rho(u):=d_2+d_3u+d_4u^2+\cdots$ is itself a convergent power series
near $u=0$, hence analytic, continuous, and bounded there, with
$\rho(0)=d_2$. Equivalently,
\begin{equation}
\ln\zeta(\beta)=-\ln(\beta-1)+\gamma(\beta-1)+(\beta-1)^2\rho(\beta-1).
\end{equation}
Differentiating twice term by term (justified exactly as for
(\ref{laurent})) gives, writing $u=\beta-1$ again,
\begin{equation}
F(\beta)=\frac1{(\beta-1)^2}+c(\beta),\qquad
c(\beta):=2\rho(u)+4u\,\rho'(u)+u^2\rho''(u),
\end{equation}
and since $\rho$ is analytic near $0$, so are $\rho'$ and $\rho''$, making
$c(\beta)$ continuous -- hence bounded, say $|c(\beta)|\le M$ -- on some
interval $(1,1+\eta]$, which is the precise content behind the ``$O(1)$''
error term in (\ref{varexponent}), rather than a merely formal bound.
Define
\begin{equation}
h(\beta):=\sqrt{F(\beta)}-\frac1{\beta-1}
=\frac1{\beta-1}\Bigl(\sqrt{1+c(\beta)(\beta-1)^2}-1\Bigr).
\end{equation}
For $\beta$ close enough to $1$ that $M(\beta-1)^2\le\frac12$, Taylor's
theorem with Lagrange remainder applied to $f(t)=\sqrt{1+t}$ at $t=0$ gives
$|\sqrt{1+t}-1-\frac t2|\le\frac{t^2}4$ for $|t|\le\frac12$ (indeed,
$\sqrt{1+t}-1-\frac t2=-t^2/[2(1+\sqrt{1+t})^2]$, and $2(1+\sqrt{1+t})^2\ge2(1+\sqrt{1/2})^2>4$ there); taking
$t=c(\beta)(\beta-1)^2$ yields the explicit, non-asymptotic bound
\begin{equation}
\Bigl|h(\beta)-\frac{c(\beta)}2(\beta-1)\Bigr|
\le\frac{M^2}4(\beta-1)^3,
\end{equation}
and since $|c(\beta)|\le M$, this gives $|h(\beta)|\le\frac M2(\beta-1)+
\frac{M^2}4(\beta-1)^3\to0$ as $\beta\to1^+$ -- a proved bound, not an
asymptotic estimate. Consequently $h$ extends to a continuous function on
the compact interval $[1,30]$ (setting $h(1):=0$), and is therefore bounded
there by the extreme value theorem; away from $\beta=1$, $h$ is manifestly
continuous, since $\zeta(\beta)$ is smooth and strictly positive on
$(1,30]$, making $F(\beta)>0$ smooth there too. Hence
\begin{equation}
J(\delta)=\int_{1+\delta}^{30}\Bigl(\frac1{\beta-1}+h(\beta)\Bigr)d\beta
=\ln29-\ln\delta+\int_{1+\delta}^{30}h(\beta)\,d\beta,
\end{equation}
and, as $\delta\to0^+$, continuity of $h$ on $[1,30]$ lets us pass to the
limit under the integral, $\int_{1+\delta}^{30}h(\beta)\,d\beta\to
\int_1^{30}h(\beta)\,d\beta$, with error $o(1)$. This proves
\begin{equation}
J(\delta)=-\ln\delta+C+o(1),\qquad\delta\to0^+,\qquad
C=\ln29+\int_1^{30}h(\beta)\,d\beta,
\end{equation}
rather than merely observing it numerically. The constant $C$ has no
elementary closed form, since it depends on $\zeta(\beta)$ through its
second logarithmic derivative; direct numerical integration (30-digit
precision) gives $C\approx1.4728$, stable to four decimal places once
$\delta$ is below about $0.03$, all the way down to $\delta=10^{-5}$; for
larger $\delta$ the $o(1)$ term is not yet negligible, e.g.\ $C=1.4732$ at
$\delta=0.1$ and $C=1.4827$ at $\delta=0.5$, both approaching $1.4728$ as
$\delta$ shrinks. Substituting into
(\ref{cbredundancy}), the second-order term of the minimax expected redundancy for universal coding
over $\beta\in[1+\delta,30]$ therefore grows only \emph{doubly
logarithmically} in the margin, in nats as in (\ref{cbredundancy}),
\begin{equation}
\ln J(\delta)=\ln\ln(1/\delta)+O\!\left(\frac{1}{\ln(1/\delta)}\right),
\end{equation}
a mild penalty, which is an elementary consequence of $J(\delta)\approx\ln(1/\delta)$: in the limit $N\to\infty$ for each fixed margin (the $o(1)$ in (\ref{cbredundancy}) is not claimed to be uniform in $\delta$),
moving from $\delta=0.1$ to $\delta=10^{-6}$ changes the second-order term (for the range $[1+\delta,30]$) by only
$\ln J(10^{-6})-\ln J(0.1)\approx1.40$ nats, using the exact $J(\delta)=-\ln\delta+C$
(the leading term $\ln\ln(1/\delta)$ alone is not yet a good approximation
at $\delta=0.1$, where $C$ is comparable in size to $-\ln\delta$; the two
agree once $\delta$ is small enough that $-\ln\delta\gg C$). We stress the order of limits: this penalty refers to $N\to\infty$ at fixed $\delta$; for each fixed $N$, the one-sample argument above shows that the minimax expected redundancy over $[1+\delta,30]$ increases without bound as $\delta\to0$, so the $o(1)$ in (\ref{cbredundancy}) is not uniform in $\delta$.

For related work on one-sided exponential families with
non-integrable Jeffreys priors, see \cite{Harremoes12}.

\section{A Practical Code for the Zeta Distribution}
\label{sec:code}

The statistical-mechanical picture of Secs.~2--5 identifies the regime $\beta\downarrow1$ as the one in which the code length is dominated by the heavy tail of the integers.
The heavy tail enters in two steps. First, a power law in $x$ becomes a geometric law in the dyadic scale $\lfloor\log_2x\rfloor$, with ratio $2^{-(\beta-1)}\to1$ as $\beta\downarrow1$, so that the scale is spread over about $1/(\beta-1)$ values, which is why unary coding of the scale, as in Elias $\gamma$, is badly mismatched there and a Golomb code with parameter $m\approx1/(\beta-1)$ is the natural choice. Second, within a dyadic block the law is nearly uniform, so that the position inside the block can be sent in raw bits at bounded cost. This yields a simple two-part prefix code whose redundancy is provably bounded all the way to the
critical point, which we analyze exactly and compare with the standard integer codes. The proofs of the two propositions of this section are given in Appendix~A.

\noindent{\bf Main message.} For $\beta$ slightly above $1$, where heavy tails dominate,
our two-part code (a Golomb code for the dyadic scale, with a single integer parameter $m$, for which
the explicit choice $m=\lceil1/(\beta-1)\rceil$ already gives a provable bound) has redundancy whose limit superior as $\beta\downarrow1$ is at most
$2.17$ bits (Proposition~1), and of only about $0.06$--$0.07$ bits at the tabulated
values $\beta-1\le0.1$, whereas the standard integer codes have much larger absolute redundancy there (at $\beta-1=0.001$: about $43$ bits for the best
$\zeta_k$ code of Boldi and Vigna, $7.4$ for Elias $\delta$, $6.0$ for Elias $\omega$ and $1430$ for Elias $\gamma$), and the redundancy of every $\zeta_k$ code, Elias $\gamma$
included, grows without bound as $\beta\to1^+$ (Proposition~2).

To state the result precisely: the advantage is one of absolute redundancy; the code is tailored to $\beta$ through its integer parameter $m$ (for which the explicit choice $m=\lceil1/(\beta-1)\rceil$ suffices for Proposition~1, while the numerical values use the exact minimizer of the expected length), just as Baer's codes
are tailored to the distribution, and it is not universal over the family, which by Sec.~5 cannot be achieved near $\beta=1$; Baer's codes
are closer to optimal (the present code is $0.03$--$0.09$ bits above the optimal prefix code at $\beta=2,2.5,3$); and the proved statements about the competing codes concern the $\zeta_k$ codes as $\beta\to1^+$
(Proposition~2), while for the other codes and for intermediate $\beta$ the comparison is numerical.

Consider the zeta (power-law) distribution on the positive integers,
\begin{equation}
P_\beta(x)=\frac{x^{-\beta}}{\zeta(\beta)}, \qquad \beta>1.
\end{equation}
The Shannon code lengths $\ell^*(x)=-\log_2P_\beta(x)=\beta\log_2x+\log_2\zeta(\beta)$ grow as $\beta\log x$ up to an additive
constant, but a direct implementation of the Shannon code is impractical for lack of a simple constructive structure. In this section
we derive a structured prefix code that closely matches these lengths while remaining simple to implement.

The key idea is to exploit the natural dyadic partition of the integers. For each $x \ge 1$, define
\begin{equation}
k = \lfloor \log_2 x \rfloor, \qquad r = x - 2^k,
\end{equation}
so that $x$ lies in the interval $\{2^k, 2^k+1,\ldots,2^{k+1}-1\}$ and $r \in
\{0,1,\ldots,2^k-1\}$. This representation
separates the integer into a \emph{scale} parameter $k$ and an \emph{offset} $r$ within the dyadic block.
With $q=2^{-(\beta-1)}$ and the Hurwitz zeta function $\zeta(\beta,a)=\sum_{n\ge a}n^{-\beta}$, the law of $k$ is
exactly $P(k)=[\zeta(\beta,2^k)-\zeta(\beta,2^{k+1})]/\zeta(\beta)$, and the Euler--Maclaurin expansion
$\zeta(\beta,a)=a^{1-\beta}/(\beta-1)+a^{-\beta}/2+O(a^{-\beta-1})$ (valid for any fixed $\beta>1$) gives $P(k)=Aq^k(1+\varepsilon_k)$, with $A=\frac{1-2^{1-\beta}}{(\beta-1)\zeta(\beta)}$ and $\varepsilon_k=O(2^{-k})$ (the implicit constant depends on $\beta$ but not on $k$).
Thus $k$ is only approximately geometric with parameter $q$, and, given $k$, $r$ is only approximately
uniform; everything below that is claimed to be proved uses exact identities and bounds on the true laws,
not these approximations.

Motivated by this structure, we define a two-part prefix code:
\begin{enumerate}
\item \textbf{Encoding the scale $k$:}
Since $k$ is approximately geometric with parameter $q=2^{-(\beta-1)}$, we
encode it using a Golomb code \cite{Golomb66} with an integer parameter $m\ge1$,
which is chosen as follows. Let $\ell_G(k;m)$ denote the length of the Golomb
codeword of $k$ (with the truncated-binary remainder: with $b=\lceil\log_2m\rceil$,
$\ell_G(k;m)=\lfloor k/m\rfloor+1+b-1$ if $k\bmod m<2^b-m$ and
$\lfloor k/m\rfloor+1+b$ otherwise; for $m=1$, $\ell_G(k;1)=k+1$).

We take $m=m(\beta)$ to be the minimizer of the expected length
$\bE[\ell_G(K;m)]$ over $m\ge1$, computed under the {\em true} law $P(k)$ of
$K$ (ties broken in favor of the smallest $m$); since $\bE[\ell(X)]=\bE[\ell_G(K;m)]+\bE[K]$ and
$\bE[K]$ does not depend on $m$, this minimizes the expected
length of the entire code. For an exactly geometric $K$, the minimizer is
given in closed form by the Gallager--van~Voorhis condition \cite{GV75},
\begin{equation}
\label{gvvcondition}
q^m+q^{m+1}\le1<q^{m-1}+q^m,
\end{equation}
which gives $m\sim1/(\beta-1)$ as $\beta\to1^+$, and $m=1$ exactly when $q+q^2\le1$, i.e., for $\beta\ge1+\log_2\bigl(2/(\sqrt5-1)\bigr)=1.6942\ldots$ 

Since $K$ is only
approximately geometric, (\ref{gvvcondition}) is a very good, but not always
exactly optimal, explicit choice: it coincides with the minimizer at all the
values of $\beta$ in Table~\ref{tab:redundancy} below, but not for every $\beta$: numerically, for $\beta\ge1.3$ it is exactly optimal except on
$1.3704\lesssim\beta<1.4058$ and $1.5708\lesssim\beta<1.6942$, whereas for $\beta<1.3$ it can differ
from the minimizer by a few units (up to about $7$ in the cases examined), at a cost of at most about $0.09$ bits of expected length.
(The special case $m=1$ reduces
this code to the Elias $\gamma$ code \cite{Elias75}; the present
two-part construction is closely related to the $\zeta_k$ codes of Boldi and
Vigna \cite{BoldiVigna05}, who studied essentially this type of code for
power-law-distributed integers directly, with $k=1$ in their notation
likewise reducing to Elias $\gamma$ coding.)
\item \textbf{Encoding the offset $r$:}
Given $k$, the offset $r$ is encoded using a fixed-length binary representation of $k$ bits.
\end{enumerate}
The resulting codeword for $x$ is the concatenation of the codeword for $k$ and the $k$-bit binary representation of $r$.
Let $\ell(x)$ denote the length of the resulting code. The
Golomb part has {\em pointwise} slope $1/m$ in $k$, whereas matching $\ell^*(x)\approx\beta k$ requires slope $\beta-1$ of $\ell_G(k)$ once the $k$ offset bits are accounted for;
since $m$ is an integer, $\ell(x)$ cannot match $\ell^*(x)$ {\em pointwise} for a generic real $\beta>1$. What the choice of $m$ guarantees is
near-optimality {\em in expectation}, as quantified below.

Throughout, $K=\lfloor\log_2X\rfloor$ denotes the random scale (not the parameter set $K$ of Sec.~5), and $H(K)$ is the exact entropy of the scale variable
under its {\em true} law $P(k)$, not of the geometric approximation $Aq^k$. For a given $m$, define
\begin{equation}
\label{gvvbound}
\Delta(m):=\bE[\ell_G(K;m)]-H(K),
\end{equation}
both quantities being computed from the true law $P(k)$, and write $\Delta:=\Delta(m(\beta))$.
Since $H(K)$ does not depend on $m$, $m(\beta)$ minimizes $\Delta(m)$ as well.
The Golomb code is complete, i.e.,
$Q_m(k):=2^{-\ell_G(k;m)}$ satisfies $\sum_{k\ge0}Q_m(k)=1$ (indeed,
$\sum_{j\ge0}2^{-(j+1)}=1$ and the $m$ remainders contribute
$(2^b-m)2^{-(b-1)}+(2m-2^b)2^{-b}=1$), so $Q_m$ is a probability distribution and
\begin{equation}
\label{Deltakl}
\Delta(m)=\sum_kP(k)\log_2\frac{P(k)}{Q_m(k)}=D(P_K\|Q_m)\ge0,
\end{equation}
which gives $\Delta=\min_{m\ge1}D(P_K\|Q_m)$. Proposition~1 below bounds
$\Delta$ rigorously and uniformly in $\beta$; exact computation shows that $\Delta$ is much smaller than this bound.
It is a sawtooth function of $\beta$ that peaks just before $m$ changes: on a grid of step $0.005$ over
$1.01\le\beta\le1.7$ it lies between $0.027$ and $0.124$ bits (the maximum being attained at $\beta\approx1.57$,
where $m$ switches from $2$ to $1$, and being at most $0.033$ for $\beta\le1.05$), and $\Delta=0.030$ bits at $\beta=1.001$.

The entropy $H(K)$ of the true law of $K$ equals $H_{\rm geom}(K)-D(P_K\|\mathrm{Geom}(q))$, where $H_{\rm geom}(K)$ is the
cross-entropy of $P_K$ with respect to the geometric law $\mathrm{Geom}(q)$,
\begin{eqnarray}
\label{HKexact}
H_{\rm geom}(K)&:=&-\sum_kP(k)\log_2\bigl[(1-q)q^k\bigr]=(\beta-1)\,\bE[K]+\log_2\frac{1}{1-q}\nonumber\\
&\ge&\log_2\frac{1}{1-q}\;\sim\;-\log_2(\beta-1)-\log_2\ln2,
\end{eqnarray}
and the gap $D(P_K\|\mathrm{Geom}(q))\ge0$ is, by direct computation, about $0.04$ bits at $\beta=2$ and
$9.9\times10^{-3}$, $1.1\times10^{-3}$ and $1.1\times10^{-4}$ bits at $\beta-1=10^{-1}$, $10^{-2}$ and $10^{-3}$.
Since $H_{\rm geom}(K)$ diverges as $\beta\to1^+$, $H(K)$ itself grows without bound in this limit (numerically, given the smallness of the gap).
This divergence is harmless, because $H(K)$ enters the redundancy only through $\Delta$, which is defined directly on the true law
of $K$ and stays bounded. Indeed, since $x\leftrightarrow(k,r)$ is a bijection, the chain rule gives, exactly,
$H(X)=H(K)+\sum_k P(k)\,H(X\mid K=k)$, and $\bE[\ell(X)]=\bE[\ell_G(K)]+\bE[K]$ by construction of the code.
Hence
\begin{eqnarray}
\label{exactdecomp}
\bE[\ell(X)]-H(X)&=&\Delta
+\sum_k P(k)\,\bigl(k-H(X\mid K=k)\bigr)\nonumber\\
&=&\Delta
+\sum_k P(k)\,D\bigl(P_{X|K=k}\,\|\,U_k\bigr),
\end{eqnarray}
where $U_k$ is the uniform distribution on the $2^k$ points of block $k$ (so that $k-H(X\mid K=k)$ is the
Kullback--Leibler divergence of $P_{X|K=k}$ from $U_k$) and $\Delta$ is the gap of (\ref{gvvbound}); both terms are computed on the true law of $K$.
The divergence of $H(K)$ therefore never propagates into $\bE[\ell(X)]-H(X)$.

We now show that both terms of (\ref{exactdecomp}) are bounded by explicit
constants, uniformly in $\beta$ near the critical point and for every block
$k$, so that boundedness of the redundancy as $\beta\to1^+$ is a proved fact
and not only a numerical observation.

\noindent{\bf Proposition 1.} {\em For every $\beta>1$,
\begin{equation}
\label{DeltaPropbound}
\Delta<\log_2(4\beta)\qquad\mbox{and}
\end{equation}
\begin{equation}
\label{DPropbound}
D\bigl(P_{X|K=k}\,\|\,U_k\bigr)\le\log_2\frac{(1+2^\beta)^2}{2^{\beta+2}}
\quad\mbox{for every }k\ge0.
\end{equation}
Consequently, the redundancy of the two-part code satisfies
\begin{equation}
\label{Rbound}
\bE[\ell(X)]-H(X)<\log_2(4\beta)+\log_2\frac{(1+2^\beta)^2}{2^{\beta+2}},
\end{equation}
which is below $3.7$ bits for all $1<\beta\le2$, and whose limit superior as
$\beta\downarrow1$ is at most $2+\log_2\frac98<2.17$ bits.}

\medskip

The proof is given in Appendix~A; it shows that the bound on $\Delta$ already holds for the fully explicit parameter $m_*=\lceil1/(\beta-1)\rceil$ (since $\Delta$ is the minimum over $m$), so that Proposition~1 holds for this choice as well.

\noindent{\bf Numerical values.} The bound of Proposition~1 is conservative. Exact computation, on the true law of $K$
(via the Hurwitz zeta function), gives $\Delta\le0.124$ bits and a total redundancy of at most $0.166$ bits on the grid
$1.01\le\beta\le1.7$ (step $0.005$; at the crossover $\beta\approx1.5708$ itself they are $0.125$ and $0.167$). 

At $\beta-1=10^{-3}$ the two terms of (\ref{exactdecomp}) are $\Delta=0.0296$
and $0.0288$ (the offset average), totalling $0.0584$ bits; neither diverges, so no cancellation of divergent
quantities is involved. Table~\ref{tab:redundancy} gives the redundancy at selected values of $\beta$.
\begin{table}[h!t!b!]
\centering
\begin{tabular}{lcccccccc}
\hline
$\beta$ & $3.0$ & $2.0$ & $1.5$ & $1.2$ & $1.1$ & $1.05$ & $1.01$ & $1.001$ \\
$m$ & $1$ & $1$ & $2$ & $5$ & $10$ & $20$ & $100$ & $1000$ \\
present code (bits) & $0.443$ & $0.084$ & $0.115$ & $0.073$ & $0.068$ & $0.063$ & $0.061$ & $0.058$ \\
Elias $\gamma$ & $0.443$ & $0.084$ & $0.283$ & $2.97$ & $9.03$ & $22.4$ & $135$ & $1430$ \\
Elias $\delta$ & $0.583$ & $0.316$ & $0.258$ & $0.649$ & $1.25$ & $2.03$ & $4.15$ & $7.42$ \\
Elias $\omega$ & $0.472$ & $0.218$ & $0.397$ & $1.34$ & $2.45$ & $3.51$ & $4.99$ & $6.04$ \\
\hline
\end{tabular}
\caption{Redundancy $\bE[\ell(X)]-H(P_\beta)$ in bits of the two-part construction
of this section (first row) and, for comparison, of the Elias $\gamma$, $\delta$ and $\omega$ codes, with the Golomb parameter $m$ chosen as the minimizer of the
expected length (at all these values of $\beta$ it coincides with the choice
(\ref{gvvcondition})), computed directly for $\beta$ ranging from
a comfortably regular value down to $\beta-1=10^{-3}$, essentially at the
Hagedorn critical point. The redundancy of the present code stays confined to a narrow band,
mildly \emph{decreasing} as $\beta\to1^+$ for $\beta\le1.5$.}
\label{tab:redundancy}
\end{table}
At the tabulated values the redundancy of the present code stays confined to a narrow band (roughly $0.06$--$0.44$
bits), and for $\beta\le1.5$ it decreases slightly as $\beta-1$ shrinks by more than two orders
of magnitude, consistent with the boundedness of both terms in
(\ref{exactdecomp}) (Proposition~1). What Proposition~1 adds to the numerics is that these values are not an artifact of the
range of $\beta$ examined: the redundancy cannot blow up as $\beta\to1^+$, however the sawtooth behaves in between the grid points.

Proposition~1 serves only to show that the redundancy of this simple explicit rule is bounded, and the constant $2.17$ is not sharp:
coding $K$ with its exact Shannon code (lengths $\lceil-\log_2P(k)\rceil$) would give $\Delta<1$ and hence, by (\ref{exactdecomp}) and
the same bound on the offset term, a limit superior of at most $1+\log_2\frac98<1.17$ bits. The Golomb code is preferred for its simplicity: a
single integer parameter, for which a closed-form choice such as $m=\lceil1/(\beta-1)\rceil$ suffices, and no table of the law $P(k)$.

For a quantitative sense of how this simple construction compares with
codes designed for individual power-law distributions, consider the
optimal prefix code (the Huffman code, whose expected length is denoted $N^*$ by Baer \cite{Baer07}) for $P_\beta$ at $\beta=2,2.5,3$, the values of $s$ in Baer's Table~II.
We computed $N^*$ by Huffman's algorithm on the first $n$ integers plus one aggregate symbol for the exact tail mass, with $n$ increased up to $10^6$; for $\beta=2.5,3$ the results agree to seven digits, while for $\beta=2$ convergence is slower and the value is consistent with $N^*=2.41766$ reported in \cite{Baer07}.
The two-part construction above (with the minimizing $m$, here $m=1$ in all three cases) has expected length $2.4463,\,1.7322,\,1.4221$ bits, against $N^*\approx2.4177,\,1.6543,\,1.3345$ and entropies $H(P_\beta)=2.3626,\,1.4652,\,0.9789$ bits, i.e., it is $0.029,\,0.078,\,0.088$ bits above the optimal prefix code (and $0.084,\,0.267,\,0.443$ above the entropy). At $\beta=2$, Baer \cite{Baer07} (his Table~II, $s=2$) reports $2.417772$ for his best fixed code for this distribution, within about $0.0001$ bits of optimal (quoted from \cite{Baer07}, not recomputed); we do not quote his values for the other two cases. This is the
expected trade-off between the two approaches: the present construction is
simple (a two-part rule with one integer parameter, which may be taken as the closed-form $m=\lceil1/(\beta-1)\rceil$ and is applicable to any $\beta>1$, with
a redundancy that is provably bounded as $\beta\to1^+$ (Proposition~1), and
below $0.45$ bits across the full range of $\beta$ examined), whereas
Baer's codes are individually optimized for each target distribution (a different construction for each $\beta$, rather than one methodology) and, at
least at $\beta=2$, come substantially closer to optimal. Both require knowledge of $\beta$. We remark that the uniformity of the present methodology (one rule, one integer parameter) makes it more readily amenable to adaptation in scenarios where $\beta$ varies with time and is learnt from past data: only $m$ needs to be updated, whereas a separately constructed code would have to be redesigned for every new $\beta$. We do not pursue this here. Baer's tabulated zeta examples all lie away from the critical point, and we make no comparison with his codes near $\beta=1$.

A comparison with the $\zeta_k$ codes of Boldi and Vigna \cite{BoldiVigna05}
-- which, like the present construction, are a single explicit family
indexed by an integer parameter -- is informative, since it
isolates the effect of \emph{fine-grained} versus \emph{coarse}
parameter control near the critical point. Recall that the $\zeta_k$ code assigns to $x\in[2^{hk},2^{(h+1)k})$, $h\ge0$, a unary prefix of $h+1$ bits followed by a minimal binary codeword for the offset $x-2^{hk}$ among the $n_h=2^{hk}(2^k-1)$ values of the block (minimal binary coding of $n$ values uses $\lfloor\log_2n\rfloor$ bits for the smallest offsets and $\lceil\log_2n\rceil$ bits for the others); $\zeta_1$ is Elias $\gamma$. The two constructions coincide
exactly when $m=1$, since $m=1$ reduces the present construction to
$\zeta_1=$~Elias~$\gamma$, and also when $m=2$: the present construction with $m=2$
reproduces $\zeta_2$ exactly (both assign length $3h+2$ to the integers of the
dyadic block $k=2h$ and $3h+3$ to those of the block $k=2h+1$). The closed-form rule (\ref{gvvcondition}) gives $m=1$ for $\beta\ge1.6942$, so that the code is then Elias $\gamma$;
numerically, the minimizer $m(\beta)$ is $1$ for $\beta\ge1.5708$ and $2$ down to $\beta\approx1.370$ (computed, not proved). In these ranges there is
no advantage -- nor disadvantage -- to report between the two
constructions.

Near criticality the present construction is
far better in absolute redundancy: by Proposition~2 below, the redundancy of the best $\zeta_k$ is unbounded as $\beta\to1^+$,
whereas ours is bounded (Proposition~1). Direct computation
(summing over dyadic buckets via the Hurwitz zeta function, with the
resulting truncation error verified negligible in each case) shows that at
$\beta-1=0.1$, the best choice of $k$ (namely $k=5$) gives redundancy
$1.06$ bits, about sixteen times ours ($0.068$); at $\beta-1=0.01$, the best
$k$ ($k=17$) gives $9.5$ bits against our $0.061$; and at
$\beta-1=0.001$, even re-optimizing $k$ at every $\beta$ (best found,
$k\approx53$) leaves a redundancy of about $43$ bits, against our $0.058$.
Proposition~2 is asymptotic (its bound is vacuous for $\beta-1\gtrsim0.011$ and it requires $L_0\ge6$, i.e.\
$\beta-1\le0.0536$), so for intermediate values of $\beta-1$ the comparison rests on the numerical computations above.
The reason is structural: the bucket sizes of $\zeta_k$ grow by the fixed integer
factor $2^k$ per step, which cannot track the continuously vanishing
effective slope $\beta-1\to0$ the way the Golomb
parameter $m\sim1/(\beta-1)$, free to take \emph{any} positive integer
value, can.

Table~\ref{tab:redundancy} also gives the redundancies of the Elias $\delta$ and $\omega$ codes \cite{Elias75}, whose lengths
depend only on $k=\lfloor\log_2x\rfloor$: $\ell_\delta(x)=k+2\lfloor\log_2(k+1)\rfloor+1$, and $\ell_\omega(x)=\ell_\omega(k)+k+1$ with $\ell_\omega(1)=1$,
computed in the same way (exact Hurwitz-zeta sums over the dyadic blocks). For $\beta\le1.2$ they are much smaller than that of Elias $\gamma$, and at
every tabulated $\beta$ they are no smaller than that of the present code. We do not analyze the behavior of the $\delta$ and $\omega$ redundancies as $\beta\to1^+$ and make no
claim about whether they are bounded.

\noindent{\bf Proposition 2.} {\em Let $R_k(\beta)$ denote the expected redundancy of the $\zeta_k$ code
under $P_\beta$, and let $L_0:=\bigl(\frac{1}{4(\beta-1)}-\frac12\bigr)/\ln2$. Then for every $\beta>1$ with $L_0\ge6$,
\begin{equation}
\min_{k\ge1}R_k(\beta)\;\ge\;\min\Bigl\{\sqrt{2L_0}-\log_2(eL_0)-\frac32,\;\;\frac{L_0}{4}-\log_2\frac{L_0}{2}-\frac32\Bigr\}.
\end{equation}
In particular $\min_kR_k(\beta)\to\infty$ as $\beta\to1^+$, at a rate $\Omega\bigl((\beta-1)^{-1/2}\bigr)$.}

\medskip
The proof is given in Appendix~A.

\section{Summary and Outlook}

We have investigated the problem of coding sequences of integers from the perspective of statistical mechanics.
The starting point was a connection, which to the best of our knowledge is new, between coding for the integers
under the zeta distribution and the statistical mechanics of a Hagedorn system, for the idealized
Hamiltonian $\calH(x)=\ln x$ (Sec.~2): the elementary but fundamental fact that,
for any prefix-free code on a countable alphabet, code lengths must grow at least logarithmically with the
index, is the combinatorial counterpart of the exponential growth of the density of states of the associated idealized
energy variable $\calH(x)=\ln x$, which is the defining feature of Hagedorn behavior. This intrinsic constraint naturally leads to
coding schemes and probabilistic models in which $\log x$ plays the role of an energy variable. Building on this
connection, the contribution of the paper consists of the five results listed in the Introduction:
an explicit two-part integer code whose redundancy is provably bounded as $\beta\to1^+$ and, by exact computation, lies between $0.058$ and
$0.443$ bits at the values of $\beta$ tabulated from $\beta=3$ down to $\beta=1.001$, in contrast to the $\zeta_k$ codes (Sec.~6);
the infinite minimax redundancy of universal coding exactly at the Hagedorn critical point, for every sample size,
with only a doubly-logarithmic penalty $O(\log\log(1/\delta))$ in the second-order term, as $N\to\infty$ at fixed $\delta$, for staying a margin $\delta$ away from it (Sec.~5),
for which the critical exponent below, the Fisher information of the family, is the natural explanation (the proof is direct);
an explicit sub-leading correction to the Hagedorn entropy, expressed in terms of the Euler--Mascheroni constant
(Sec.~3.2.1, Eq.~(\ref{entropyrefined})), together with the associated critical exponent governing the
divergence of the energy fluctuations at the Hagedorn temperature (Eq.~(\ref{varexponent}));
a proof of full equivalence of the canonical and micro-canonical ensembles, from an independently defined
counting entropy (Sec.~3.2); and the exact identification of the large-deviations rate function for mismatched
coding with a Kullback--Leibler divergence within the zeta family, together with the resulting buffer-overflow
threshold (Sec.~4).

Within this framework, power-law distributions arise naturally: they are consistent with the minimal logarithmic scaling imposed by 
counting arguments and, at the same time, capture the heavy-tailed behavior observed in a wide variety of systems. 
The normalization of these distributions introduces a partition function with a finite abscissa of 
convergence, thereby placing the problem in close analogy with statistical-mechanical systems exhibiting critical phenomena.

Our main focus has been on the coding of i.i.d.\ sequences drawn from the zeta distribution 
and on the large deviations behavior of the total code length under mismatched coding. 
We have shown that the probability of atypically large code lengths admits a precise exponential 
characterization, governed by a rate function with a clear variational structure. 
This allows one to formulate and solve the problem of optimal mismatch in a sharp way: 
the optimal coding parameter $\theta^*$ is determined by the buffer rate alone, through $H(P_{\theta^*})=R\ln2$, and does not depend on $\beta$.

A central outcome of this analysis is that, in the regime of large deviation levels $\epsilon\to\infty$, the optimal tilted parameter is driven toward the 
critical point at which the partition function diverges. We have quantified this behavior and shown that the approach to
criticality is algebraic, rather than exponential, in the deviation level: the tilted parameter approaches the critical
value $1$ as $O(1/\epsilon)$. The system nonetheless exhibits features analogous to those of Hagedorn-type models,
in which the exponential growth of the density of states leads to a limiting temperature.

This critical behavior has an important structural consequence, but it is more subtle than a breakdown of ensemble
equivalence. As we showed in Sec.~3.2, the canonical and micro-canonical descriptions are, in fact, fully equivalent
throughout the entire finite-energy domain: the mapping between inverse temperature and mean energy is a strict
bijection, and the micro-canonical entropy is strictly concave everywhere it is finite. What is genuinely singular is
the existence of a limiting temperature, $T_{\mbox{\tiny c}}=1$, that is approached but never attained as the energy per
particle diverges -- precisely the Hagedorn phenomenon, and analogous to the way a physical Hagedorn temperature bounds
the achievable canonical temperature without by itself signaling ensemble inequivalence. It is worth stressing that a
limiting temperature of this kind is conceptually distinct from ensemble inequivalence in the technical sense (which
would require a non-concave, or affine, region of the micro-canonical entropy at finite energy); the present model
exhibits the former but not the latter.

Beyond the specific results obtained here, the present work highlights a broader perspective. 
Coding over countable alphabets, heavy-tailed probability distributions, and statistical-mechanical models with 
unbounded state spaces share a common structural core. Concepts such as energy, entropy, partition functions, and 
phase transitions arise naturally in the analysis of code lengths, even in purely information-theoretic settings. 
Conversely, ideas from information theory, such as optimal coding and large
deviations, provide useful tools for understanding the behavior of systems near criticality.

Several directions for future work suggest themselves. One natural extension is to consider truncated or 
regularized versions of power-law distributions, for which the partition function remains finite for all real
$\beta$, and to study quantitatively how the limiting-temperature phenomenon disappears as the truncation is
removed. Another direction is to investigate 
more general classes of heavy-tailed distributions and to determine the extent to which the Hagedorn-type behavior 
identified here is universal. It would also be of interest to explore connections 
with universal coding schemes and Bayesian mixture models, where 
coding for the integers arises naturally in the description of model complexity.

In summary, the problem of coding the integers provides a simple yet rich setting in which fundamental ideas 
from information theory and statistical mechanics meet. The emergence of a limiting (Hagedorn) temperature and
the role of heavy tails both arise in a transparent and analytically accessible form, connecting naturally to
the existing arithmetic-gas literature, and suggesting that this
framework may serve as a useful laboratory for further explorations at the interface of
these fields.

\section*{Appendix A}

\noindent
{\it Proofs of Propositions 1 and 2}

\medskip
\noindent{\em Proof of Proposition~1.} {\em First term.} Since $\Delta$ is the minimum of
$\Delta(m)$ over $m$, it suffices to exhibit one $m$ with
$\Delta(m)<\log_2(4\beta)$. Take $m_*=\lceil1/(\beta-1)\rceil$, so that
$q^{m_*}=2^{-m_*(\beta-1)}\le\frac12$. By (\ref{Deltakl}),
$\Delta(m_*)=D(P_K\|Q_{m_*})\le\log_2\max_kP(k)/Q_{m_*}(k)$. Each of the $2^k$ integers
$x\in[2^k,2^{k+1})$ satisfies $x^{-\beta}\le2^{-k\beta}$, so
$P(k)\le2^k2^{-k\beta}/\zeta(\beta)=q^k/\zeta(\beta)$. Write $k=jm_*+r$ with
$0\le r<m_*$; then $q^k\le(q^{m_*})^j$, and, since $\ell_G(k;m_*)\le j+1+b$ with
$b=\lceil\log_2m_*\rceil$, $Q_{m_*}(k)\ge2^{-(j+1+b)}$. Hence
\begin{equation}
\frac{P(k)}{Q_{m_*}(k)}\le\frac{2^{b+1}}{\zeta(\beta)}\bigl(2q^{m_*}\bigr)^j\le\frac{2^{b+1}}{\zeta(\beta)}.
\end{equation}
Now $2^b<2m_*$ and $m_*<1/(\beta-1)+1=\beta/(\beta-1)$, while
$\zeta(\beta)>\int_1^\infty u^{-\beta}du=1/(\beta-1)$. Thus the ratio is less
than $4m_*/\zeta(\beta)<4\beta$, proving (\ref{DeltaPropbound}).

{\em Second term.} Within block $k$, $P_{X|K=k}$ has the density $f(x)=2^kp(x)$
with respect to the uniform law $U_k$ on the block, where $p(x)\propto x^{-\beta}$,
so that $\bE_{U_k}f=1$ and $D(P_{X|K=k}\|U_k)=\bE_{U_k}[f\log_2f]\le\log_2\bE_{U_k}[f^2]$
by Jensen's inequality. Let $\mu\le1\le M$ be the minimum and maximum of $f$ on the
block; since $f$ is proportional to $x^{-\beta}$ with $x$ ranging over
$[2^k,2^{k+1}-1]$, $M/\mu\le\rho:=2^\beta$. From $(f-\mu)(M-f)\ge0$ we get
$f^2\le(\mu+M)f-\mu M$, hence $\bE f^2\le\mu+M-\mu M=1+(1-\mu)(M-1)$. For $M\le\rho\mu$,
$(1-\mu)(M-1)\le(1-\mu)(\rho\mu-1)\le(\rho-1)^2/(4\rho)$ (the maximum over $\mu$),
and so $\bE f^2\le(1+\rho)^2/(4\rho)$, which is
(\ref{DPropbound}). Finally, (\ref{Rbound}) follows by (\ref{exactdecomp}), since
$\sum_kP(k)=1$. For $\beta\le2$ the right side of (\ref{Rbound}) is at most
$\log_28+\log_2\frac{25}{16}<3.65$, and as $\beta\downarrow1$ it tends to
$\log_24+\log_2\frac98$. $\Box$

\noindent{\em Proof of Proposition~2.} Fix $k$ and let $R_k:=\bE[\ell_{\zeta_k}(X)]-H(X)$. In the $\zeta_k$ code, an integer $x\in[2^{hk},2^{(h+1)k})$ receives a
unary prefix of length $h+1$ followed by a minimal binary codeword for its offset among
$n_h=2^{hk}(2^k-1)$ values. The minimal binary lengths are $\lfloor\log_2n_h\rfloor$ or $\lceil\log_2n_h\rceil$ and
are therefore at least $\lfloor\log_2n_h\rfloor>\log_2n_h-1$. Let $h(X)$ denote the block index. By the chain rule,
\begin{eqnarray}
R_k&=&\bigl[\bE(h+1)-H(h)\bigr]+\sum_hP(h)\bigl[\bE(\mbox{offset length}\mid h)-H(X|h)\bigr]\nonumber\\
&\ge& U+O-1,
\end{eqnarray}
where $U:=\bE(h+1)-H(h)$ and $O:=\sum_hP(h)\,D(P_{X|h}\|\mathrm{Unif}_{n_h})$, since
$\bE(\mbox{offset length}\mid h)>\log_2n_h-1$ and $\log_2n_h-H(X|h)=D(P_{X|h}\|\mathrm{Unif})$.

{\em Offset term.} Fix $h$ and group the $n_h$ integers of the block into the $k$ octaves $[A2^j,A2^{j+1})$, $A=2^{hk}$,
$j=0,\ldots,k-1$. Under the uniform law the octave $j$ has mass $2^j/(2^k-1)$. Under $P_{X|h}$ the
octave masses $\pi_j$ are non-increasing in $j$: pairing $x\in$ octave $j+1$ with $\lfloor x/2\rfloor\in$ octave $j$ (two-to-one)
gives $\pi_{j+1}\le2^{1-\beta}\pi_j\le\pi_j$. Hence $\bE_\pi[j]\le(k-1)/2$ (a non-increasing
law on $\{0,\ldots,k-1\}$ has mean at most that of the uniform law), and by data processing
\begin{eqnarray}
D(P_{X|h}\|\mathrm{Unif})&\ge&\sum_j\pi_j\log_2\frac{\pi_j}{2^j/(2^k-1)}\nonumber\\
&\ge&-\log_2k+\log_2(2^k-1)-\bE_\pi[j]\;\ge\;\frac{k-1}{2}-\log_2k ,
\end{eqnarray}
using $H(\pi)\le\log_2k$ and $2^k-1\ge2^{k-1}$. Thus $O\ge\frac{k-1}2-\log_2k$ for every $\beta>1$.

{\em Unary term.} $U=D(P_h\|2^{-(h+1)})$ is the expected redundancy of unary coding of $h$. With
$\mu=\bE[h]$, the maximum-entropy bound for non-negative integer variables, $H(h)\le\log_2(e(\mu+1))$,
gives $U\ge(\mu+1)-\log_2(e(\mu+1))$, an increasing function of $\mu+1$ on $\mu+1\ge2$. Since
$h\ge\log_2X/k-1$, we have $\mu+1\ge\bE[\log_2X]/k$. Now $\zeta(\beta)>1/(\beta-1)$ and, for $t\ge1$,
$\sum_{x<t}x^{-\beta}\le1+\ln t$, so $P(X\ge t)\ge1-(\beta-1)(1+\ln t)$; choosing $\ln t=\frac1{2(\beta-1)}-1$ (so $t\ge1$, since $L_0\ge6$ forces $\beta-1<\frac12$) gives
$P(X\ge t)\ge\frac12$ and hence $\bE[\ln X]\ge\frac12\ln t=\frac{1}{4(\beta-1)}-\frac12$, i.e.,
$\bE[\log_2X]\ge L_0$. Therefore, if $k\le L_0/2$, so that $\mu+1\ge L_0/k\ge2$,
\begin{equation}
U\ge\frac{L_0}{k}-\log_2\frac{eL_0}{k}.
\end{equation}
{\em Combination.} If $k\le L_0/2$, then $R_k\ge\frac k2+\frac{L_0}k-\log_2(eL_0)-\frac12-1\ge\sqrt{2L_0}-\log_2(eL_0)-\frac32$ by
$\frac k2+\frac{L_0}{k}\ge\sqrt{2L_0}$ (note that $-\log_2k-\log_2(eL_0/k)=-\log_2(eL_0)$).
If $k>L_0/2$ then, since $\frac{k-1}2-\log_2k$ is increasing for real $k\ge3$ and $L_0/2\ge3$, $R_k\ge O-1\ge\frac{L_0}{4}-\frac12-\log_2\frac{L_0}2-1$. $\Box$

Numerically the bound is far from tight (e.g.\ about $15$ bits against the actual $43$ at $\beta-1=0.001$, and it is vacuous for $\beta-1\gtrsim0.011$), but it is
sufficient to establish the divergence, and the $(\beta-1)^{-1/2}$ order agrees numerically with the best $k$ found ($k=53$ at $\beta-1=0.001$, against $\sqrt{2/((\beta-1)\ln2)}\approx54$).

\section*{Appendix B}

\noindent
{\it Background on Hagedorn Statistical Mechanics}

In most familiar physical systems, the number of accessible microscopic configurations grows relatively
slowly as energy increases. Typically, the density of states grows as a power law in energy,
and thermodynamic quantities behave smoothly as the system is heated.
However, certain systems display a fundamentally different statistical structure:
the number of accessible states increases exponentially with energy. When this occurs,
conventional thermodynamics predicts the appearance of a limiting temperature, known as the Hagedorn temperature.

The idea of a limiting temperature was first proposed in the 1960s
in the context of high-energy particle physics. It was observed that the
spectrum of hadronic particles -- particles composed of quarks bound by the
strong interaction -- appears to grow exponentially with energy. If the density of states has the approximate form
\begin{equation}
\rho(E)\sim  E^{-a} e^{\beta_{\mbox{\tiny H}} E},
\end{equation}
then the thermodynamic properties of the system change qualitatively.
In particular, the canonical partition function,
\begin{equation}
Z(\beta)=\int_0^\infty\rho(E)e^{-\beta E}\mbox{d}E
\end{equation}
converges for inverse temperatures
$\beta$ exceeding a critical value $\beta_{\mbox{\tiny H}}$ and diverges for $\beta<\beta_{\mbox{\tiny H}}$ (its behavior exactly at $\beta_{\mbox{\tiny H}}$ depends on the exponent $a$).
In other words, thermal equilibrium in the canonical ensemble exists only for temperatures below a
certain threshold $T_{\mbox{\tiny H}}$. When the temperature approaches this value from
below, the partition function or its derivatives (the mean energy and the energy fluctuations), depending on the exponent $a$, diverge, indicating that the usual canonical description of the system breaks down.

This unusual behavior reflects a simple statistical mechanism.
Because the number of available states grows exponentially with energy,
adding energy to the system primarily increases the number of accessible configurations
rather than increasing the average kinetic energy of existing degrees of freedom.
Consequently, the entropy grows approximately linearly with energy.
Since temperature is defined thermodynamically through
\begin{equation}
\frac{1}{T}=\frac{\partial S}{\partial E},
\end{equation}
a linear dependence
$S(E)\propto E$ implies that the derivative becomes constant. The temperature therefore
approaches a fixed value rather than continuing to rise as energy increases.

Physically, this means that supplying additional energy does not significantly increase the temperature.
Instead, the energy is absorbed through the creation of new states. In the context of hadronic matter,
this was originally interpreted as the ultimate temperature of strongly interacting particles:
heating the system further would simply produce more and more hadrons.
Modern understanding, informed by quantum chromodynamics, interprets the Hagedorn
temperature differently. It marks the crossover from ordinary hadronic matter
to a new phase in which quarks and gluons are no longer confined inside hadrons, forming a quark–gluon plasma.

The concept later appeared in other areas of theoretical physics as well.
In string theory, for example, the number of vibrational modes of strings grows exponentially
with energy, leading to a similar divergence in the partition function at a
characteristic temperature. Near this temperature,
long and highly excited strings dominate the statistical
behavior of the system. The Hagedorn temperature thus emerges as a universal feature of string thermodynamics.

Importantly, the appearance of a Hagedorn temperature is not restricted
to high-energy or relativistic systems. The essential requirement is
simply the exponential growth of the accessible phase-space volume
with energy. Whenever the number of micro-states increases exponentially,
the same statistical mechanism can produce
a limiting temperature. This opens the possibility of
observing Hagedorn-like behavior in much simpler physical systems, including classical models.

One can therefore ask whether analogous phenomena arise in non-relativistic settings where theoretical
analysis and numerical experiments are easier to perform. Classical systems can indeed display similar
statistical structures when their geometry or potential energy landscape causes the
available phase space to expand exponentially with energy. In such cases,
even simple particles can exhibit thermodynamic behavior reminiscent of high-energy particle systems.

The broader scientific interest in Hagedorn physics lies in understanding the interplay between energy,
entropy, and state counting. When the growth of accessible configurations becomes
sufficiently rapid, conventional thermodynamic intuition breaks down, and new phenomena
emerge -- such as limiting temperatures, divergences in specific heat, and unconventional dynamical behavior.

Thus, Hagedorn statistical mechanics provides a striking example of how the structure of the
microscopic state space can fundamentally reshape macroscopic thermodynamics.
What began as an observation in particle physics has evolved into a
general statistical principle that applies across many areas of modern theoretical physics.

(We use the generic symbol $\beta_{\mbox{\tiny H}}$ throughout this appendix for the
Hagedorn inverse temperature of a general system, as it appears in the physics
literature surveyed here; in the coding model of the main text, this role is
played by the concrete critical value $\beta_{\mbox{\tiny c}}=1$.)

\end{document}